\documentclass[reprint,amsmath,amssymb,aps,pre,superscriptaddress,floatfix]{revtex4-1}

\usepackage{dcolumn}
\usepackage{bm}

\usepackage{amsmath}
\usepackage{color}
\usepackage[usenames,dvipsnames,svgnames,table]{xcolor}
\usepackage[table]{xcolor}
\usepackage[english]{babel}
\usepackage{graphicx}
\usepackage{caption}
\usepackage{subcaption}
\usepackage{amsmath}
\usepackage{array}
\usepackage{multirow}
\usepackage{natbib}

\newcommand{\argmax}{\operatorname*{arg\,max}}
\newcommand{\argmin}{\operatorname*{arg\,min}}

\newcommand{\blue}[1]{{#1}}
\newcommand{\HB}[2]{{#1--H$\cdots$#2}}
\newcommand{\onecol}{0.95\columnwidth}

\begin{document}
\title{Recognizing molecular patterns by machine learning: \\an agnostic structural definition of the hydrogen bond.}

\author{Piero Gasparotto}
\affiliation{Laboratory of Computational Science and Modeling, and National Center for Computational Design and Discovery
of Novel Materials MARVEL, IMX, {\'E}cole Polytechnique F{\'e}d{\'e}rale de Lausanne, 1015 Lausanne, Switzerland}

\author{Michele Ceriotti}
\email{michele.ceriotti@epfl.ch}
\affiliation{Laboratory of Computational Science and Modeling, and National Center for Computational Design and Discovery
of Novel Materials MARVEL, IMX, {\'E}cole Polytechnique F{\'e}d{\'e}rale de Lausanne, 1015 Lausanne, Switzerland}

\date{\today}


\begin{abstract}
The concept of chemical bonding can ultimately be seen as a rationalization of the
recurring structural patterns observed in molecules and solids. Chemical intuition
is nothing but the ability to recognize and predict such patterns, and how they 
transform into one another. Here we discuss how to use a computer to identify
atomic patterns automatically, so as to provide an algorithmic definition of a bond based
solely on structural information. We concentrate in particular on hydrogen bonding
-- a central concept to our understanding of the physical chemistry of water, biological
systems and many technologically important materials. Since the hydrogen bond is a somewhat 
fuzzy entity that covers a broad range of energies and distances, many different criteria
have been proposed and used over the years, based either on sophisticate electronic 
structure calculations followed by an energy decomposition analysis, or on
somewhat arbitrary choices of a range of structural parameters that is deemed
to correspond to a hydrogen-bonded configuration.
We introduce here a definition that is univocal, unbiased, and adaptive, based
on our machine-learning analysis of an atomistic simulation. The strategy we propose 
could be easily adapted to similar scenarios, where one has to recognize or 
classify structural patterns in a material or chemical compound.
\end{abstract}

\maketitle

\section*{Introduction}

Modern chemistry can be regarded as the (very successful) effort of rationalizing the 
stability and reactivity of compounds in terms of chemical bonding.\cite{paul1960book} A theory that aims
at predicting the structure and formation of chemical bonds from first principles cannot 
disregard a more or less explicit quantum mechanical description of electrons. 
However, from a purely heuristic point of view, one could also interpret the 
recurrence of structural patterns, in terms of distances and angles between atomic nuclei,
as a sign of the underlying chemical bond. In fact, chemical intuition builds on 
the possibility of recognizing these patterns and predicting whether and how they
can be modified during a chemical reaction. 

Electronic structure methods -- from density-functional theory (DFT) and Hartree-Fock to 
more accurate and demanding quantum chemistry approaches -- make it possible to
predict with great accuracy the structure and the stability of a molecule or a material
from first principles,\cite{car-parr85prl,scha-henr12book} and empirical force fields have been parametrized to reproduce 
these results at a fraction of the cost.\cite{mack+98JPhysChemB,behl-parr07prl,bart+10prl}
A computer simulation based on these techniques makes it possible to observe atomic configurations 
that are consistent with their energetic stability and the thermodynamic conditions, so in many cases the problem
one faces in computational modeling is not so much to \emph{predict} recurring 
structural patterns, as to \emph{recognize} them in a clear, unbiased way. 

An excellent example of this kind of scenario is that of recognizing hydrogen bonds. 
Since it was made popular by Pauling,\cite{paul1928PNAS} the concept of hydrogen bond has been a cornerstone in our
understanding of water, biological systems and supramolecular aggregates.\cite{paul51PNAS,hong-karp94JPhysChem,jeff97book} Spanning a range
of bond energies going from almost covalent bonds to Van Der Waals interactions,
hydrogen bonds can form and break easily at room temperature, explaining their 
importance for biological processes.\cite{desi-stei01book} While in most cases the hydrogen bond
can be understood as an electrostatic effect -- and simple empirical models based
on point charges are capable of qualitatively reproducing the structure and 
energetics obtained with electronic structure methods -- there are also
examples in which one can recognize a strongly covalent character, which is often
associated with quantum delocalization of the proton along the bond.\cite{lin+11jsp,mcke+14jcp} 
Because of their flexibility, hydrogen bonds escape an obvious, universal 
description, and several definitions exist based on somewhat arbitrary 
ranges of structural parameters\cite{wein-klei2012MolPhys} and/or decompositions of the total energy
obtained from electronic structure calculations.\cite{wend+10jpca,azarhead12JChemPhys,wein-klei14ChemEducResPract,kuhn-khal14jacs}

This paper aims at introducing a general protocol to automatically analyze 
the outcome of an atomistic simulation, in order to recognize recurring structural patterns
in a quantitative, deterministic and unbiased way. Starting from a (possibly) 
high-dimensional description of groups of atoms, for instance the set of inter-atomic 
distances, we propose to infer a Gaussian mixture model for the probability 
distribution of these initial coordinates. In order to obtain a deterministic,
parameter-less partitioning of the probability density it is best to first 
apply a mode-seeking algorithm\cite{veda-soat08chapterbook} to perform a non-parametric clustering, 
and then to find the best Gaussian fit to each mode of the distribution. 
One (or more) of the clusters can then be traced to the structural pattern(s)
of interest. What is more, this approach gives a natural out-of-sample 
probabilistic definition of a structural pattern, that can be used to identify the same 
patterns in new configurations of the system.

We dub this algorithm PAMM (probabilistic analysis of molecular motifs),
and describe in detail its rationale, functioning and implementation
in Section~\ref{sec:methods}.  We will then focus on the hydrogen bond as an 
instructive and challenging benchmark of our procedure, present several different
examples in Section~\ref{sec:results}, and finally give our conclusions.

\section{A probabilistic analysis of molecular motifs \label{sec:methods}}

The approach we introduce aims at using machine-learning algorithms to assist the 
interpretation of a computer experiment, and its rationalization in terms
of recurring structural patterns and different forms of chemical bonding, automatically
extracting  information from the large data sets that are produced by 
atomistic simulations. We focus on using exclusively structural information
to guide the analysis, partly because this is just more convenient than
performing an energy decomposition analysis on top of an electronic 
structure calculation, partly because the (free) energetic stability of different 
molecular patterns is implicit in the frequency with which they appear in 
the simulation. \blue{We will formulate the discussion of the algorithm in very general
terms, but use as a concrete example the case of the hydrogen bond in water,
that we will discuss in more detail in Section~\ref{sec:results}.}

The analysis we propose is based on a preliminary atomistic simulation that 
produces configurations consistent with the energetics of the system of interest,
and consists in three consecutive steps: 
\begin{enumerate}
  \item \textit{Definition of the training set}: after having performed a reference simulation,
  one has to introduce a (possibly high-dimensional) description of the groups of atoms that might
  be involved in recurring molecular patterns. 
  This step yields a set $\mathcal{X}=\left\{\mathbf{x}_i\right\}$ \blue{that contains $N$ vectors
  of dimensionality $D$}, that represent molecular configurations observed in the simulation. 
  A kernel density estimation 
  is then used to evaluate the probability density that underlies the distribution of
  $\mathcal{X}$.
  \item \textit{Determination of the mixture model}: the probability density of $\mathcal{X}$ is 
  analyzed to recognize the different modes of the distribution, and to 
  partition $\mathcal{X}$ in $n$ disjoint clusters \blue{using the quick-shift algorithm~\cite{veda-soat08chapterbook}.} 
  These clusters are then used to build a Gaussian mixture model, giving a probabilistic framework
  to associate regions of the $D$-dimensional space to one or more recurring patterns. 
  \item \textit{Analysis of the simulation}: the mixture model can then be used to give qualitative and 
  quantitative insight on the system being studied, and possibly as the basis for defining
  more complex orders parameters to describe and bias collective rearrangements of the various
  molecular patterns. 
\end{enumerate}

A simple library to apply PAMM to an arbitrary data set, and in the specific case of hydrogen-bond
recognition is briefly discussed in Appendix~\ref{app:pamm}, and provided as open-source code
in the Supplementary Materials (SM)\cite{si}.

\begin{figure}[tbhp]
    {\centering \includegraphics[width=\onecol]{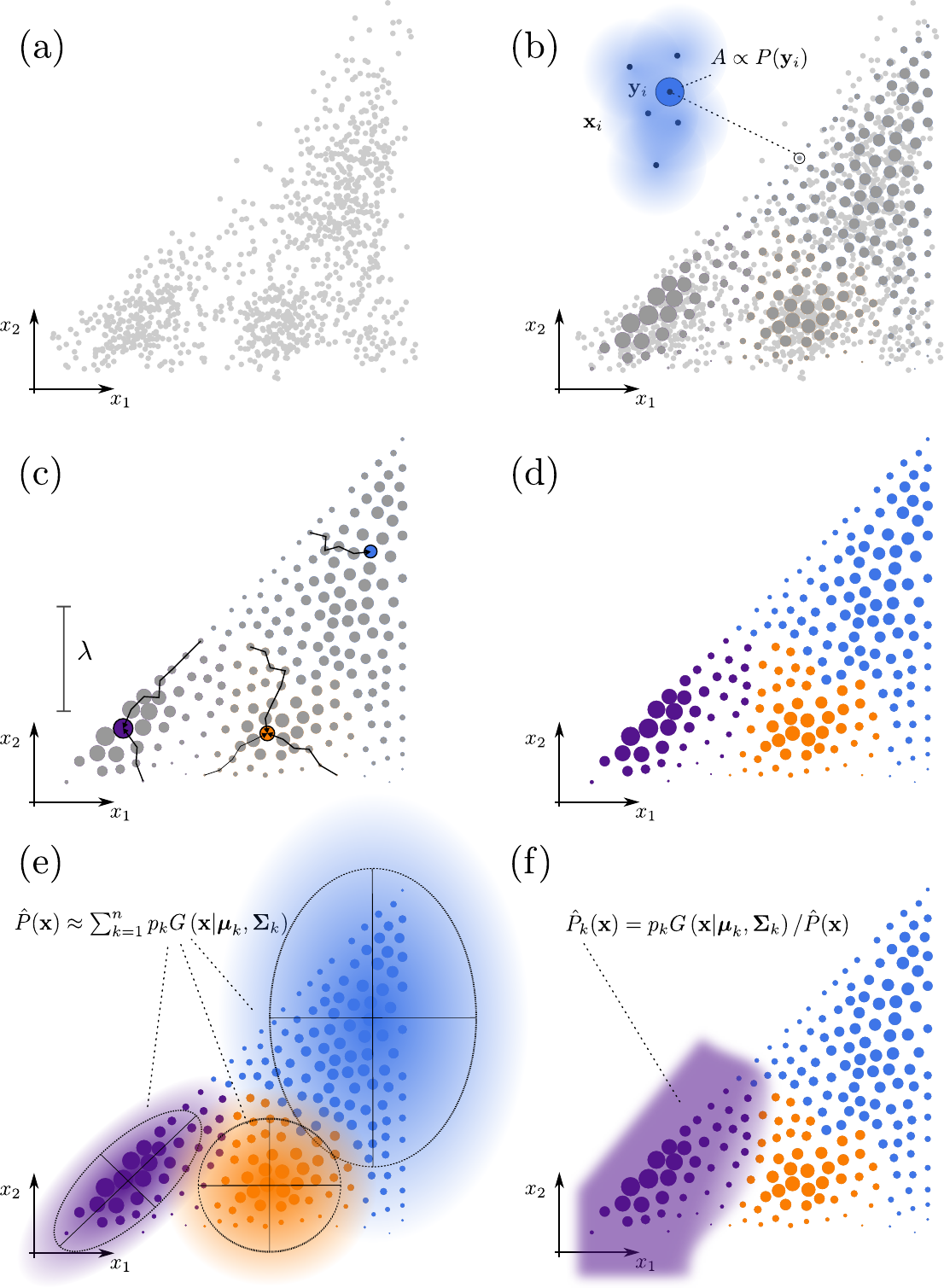} }  
    \caption{\label{fig:pamm} \blue{Simplified representation of the application of PAMM to a simple 
    two-dimensional dataset. (a) A series of data points giving a (possibly high-dimensional)
    description of molecular patterns is taken as the input. (b) A sparse grid of points is selected
    by a \emph{minmax} algorithm, and a kernel-density estimation of the probability distribution
    is evaluated at each point. (c) A \emph{quick-shift} procedure is performed to assign each of 
    the grid points to one of the modes of the probability distribution. (d) Clusters are determined
    based on the assignment to probability maxima. (e) A Gaussian mixture model is built by fitting
    separately a Gaussian to each cluster. (f) The posterior probabilities of the Gaussian mixture 
    model provide a natural, fuzzy definition of regions of parameter space that define recurring 
    molecular patterns.} }
\end{figure}

\subsection*{Definition of the training set}

The analysis we propose starts by introducing descriptors of the relative arrangement of groups
of atoms, for which one wishes to identify recurring patterns. If one wanted to recognize the existence
of a bond between two atomic species, for instance, one could process the configurations from an atomistic
simulation to output the list of distances between the pairs of atoms of the two species. 
\blue{For more complex structural patterns, one could pick all the possible tuples of atoms of a few selected 
species, and describe them in terms of all the pair-wise distances among them, 
possibly sorting groups of distances to account for the permutation of identical atoms~\cite{gall-piet13jcp}}.
In general, one would obtain a list of $N$, $D$-dimensional vectors 
$\mathcal{X}=\left\{\mathbf{x}_i\right\}$ (Figure~\ref{fig:pamm}(a)) that contains the descriptors 
for all of the tuples of selected atom kinds that are found in the simulation \footnote{In practice, one can almost
invariably introduce a cutoff to avoid considering tuples that involve atoms that are very far away
from each other, and hence clearly unrelated.}. 

\blue{
For example, a putative hydrogen bond in water involves one hydrogen atom H and a donor oxygen atom O
and an acceptor O$'$. The three distances between these atoms suffice to determine completely the geometry
of the group. The training data set $\mathcal{X}$ could then be composed of 3-dimensional vectors
$x_i=\left(\nu=d(\text{O}_i-\text{H}_i)-d(\text{O}'_i-\text{H}_i),\right.$ $\mu=d(\text{O}_i-\text{H}_i)+d(\text{O}'_i-\text{H}_i),$
$\left.r=d(\text{O}_i-\text{O}'_i)\right)$, each describing the configuration of a triplet of atoms that 
could possibly form a hydrogen bond.  }

One can then use the data set $\mathcal{X}$ to evaluate an estimate of the 
underlying probability distribution $P\left(\mathbf{x}\right)$. 
Since the $N$ points are distributed irregularly, and
the possibly very high dimensionality of the data set makes it impractical to define
the probability density on a regular grid, we used kernel density estimation (KDE) \cite{silv86book} to 
compute an estimate of  $P\left(\mathbf{x}\right)$. 

For computational efficiency, we first selected a sub-set of 
the data samples $\mathcal{Y} \subseteq \mathcal{X}$ using a \emph{minmax} criterion.\cite{elda+97ieee} 
We chose a first point $\mathbf{y}_1\in \mathcal{X}$ at random, and then iterated
\begin{equation}
\mathbf{y}_{j+1} = \argmax_{\mathbf{y}\in\mathcal{X}} \left[ \min_{i\le j} \left|\mathbf{y}_{i}-\mathbf{y}\right|\right].
\end{equation}  
Each new point is the sample with the maximal minimum distance to the points that had already
been selected. The procedure was repeated until $M$ points had been chosen, forming a sparse 
grid on which the probability is to be estimated (Figure~\ref{fig:pamm}(b)). We found that $M=\sqrt{N}$ gives a good compromise
between the density of the $\mathcal{Y}$ grid and the computational cost of the procedure 
\blue{(see the supplementary information for a discussion of the 
sensitivity of the method to this and other parameters).}

The kernel density estimate on each grid point is defined as 
\begin{equation}
   P(\mathbf{y}_i)=\sum_{j=1}^{N} w_j K\left(\left|\mathbf{x}_j-\mathbf{y}_i\right|, \sigma_{j} \right)/\sum_{j=1}^{N} w_j,
   \label{eq:prob-kde}
\end{equation}
where we used a Gaussian kernel
\begin{equation}
   K(x,\sigma)=\left(2\pi\sigma^2\right)^{-D/2}e^{-\frac{x^2}{2\sigma^2}}
\end{equation} 
and we introduced for each sample a weight $w_j$ and an adaptive kernel width $\sigma_j$. 
The weights can be for instance defined to compensate for the trivial dependence of the
probability density on the phase space volume due to the definition of the tuples, 
i.e. $w_j=w(\mathbf{x}_j)$, where the weight function is defined so that a random 
distribution of atoms would give a constant $P(\mathbf{x})$\footnote{As a 
trivial example of such weights, consider the  $1/r^2$ normalization that is used
to define the radial distribution function between a pair of atoms
based on the histogram of the pairwise distances $r$.}.

The method does not depend dramatically on the choice of the kernel width, but we
found it convenient to define automatically an adaptive width as follows. For each of 
the grid points we evaluated the minimum distance of the surrounding grid points,
$\delta_i = \min_{j\ne i} \left|\mathbf{y}_j-\mathbf{y}_i\right|$. We then associated
each of the sample points $\mathbf{x}_j$ to the nearest grid point $\mathbf{y}_i$,
and set $\sigma_j=\delta_i$. Associating each sample point to the nearest grid point
is also useful to speed up the evaluation of Eq.~\eqref{eq:prob-kde}, as one can
restrict the evaluation of the density kernel to points that are associated with 
the grid sites within a reasonable cut-off distance.

\subsection*{Determination of the mixture model}

Having computed the kernel density estimation of $P(\mathbf{y}_i)$, 
we could then proceed to \blue{subdivide it in distinct clusters, that
will be associated to one or more recurring molecular patterns}.

In order to fulfill our goal of having a univocal, deterministic pattern recognition,
we opted for a non-parametric clustering of $P(\mathbf{x})$ based on mean shift~\cite{coma-meer02ieee},
a method that identifies the modes of the probability distribution and subdivides sample 
points by following steepest ascent paths starting from each point and clustering together
the paths that converge onto the same local probability maximum. 
\blue{This approach has a profound physical interpretation: each probability maximum corresponds 
to a minimum in the free energy associated with the $D$-dimensional description of the group of atoms.
Each of the clusters then corresponds to a meta-stable configuration.}
                                   
For this application, we found 
that the most efficient and stable variant of the mean shift idea was the 
\emph{quick shift} algorithm~\cite{veda-soat08chapterbook}.
Given a set of data points and an estimate of the density
at each point ($\mathbf{y}_i$ and $P(\mathbf{y}_i)$ in our case), quick shift builds a 
tree in which each data point is a node, and the root is the point with the highest density. 
\blue{Starting from each point, one connects it to}
the nearest point that has larger probability density,
i.e. $\mathbf{y}_i$ is connected with the $\mathbf{y}_j$ such that
\begin{equation}
j=\argmin_{P(\mathbf{y}_j)>P(\mathbf{y}_i)} \left|\mathbf{y}_i-\mathbf{y}_j\right|  
\label{eq:quick-shift}
\end{equation}
The procedure defined by Eq.~\eqref{eq:quick-shift} will always end at the grid
point with the highest associated probability density. In order to obtain individual modes
of $P(\mathbf{x})$, one can introduce a length scale $\lambda$, and stop moving 
to points with higher $P$ whenever one cannot find one within $\lambda$ of the current
point.\footnote{Quick shift is not particularly sensitive to the choice of $\lambda$
We found it convenient to set $\lambda=5\left<\sigma_i\right>$, where the 
$\sigma_i$ are the kernel widths associated with the grid points. The SM~\cite{si}.}. 
By setting a cutoff for the step length, quick shift identifies a set of $n$ 
local maxima $\mathbf{z}_k$, and partitions $\mathcal{Y}$ into 
$n$ disjoint sets $\mathcal{Z}_k$ based on the maximum to which a quick shift procedure started at
each $\mathbf{y}_i$ converges to (Figure~\ref{fig:pamm}(c) and (d)). 

One could then use the sets $\mathcal{Z}_k$ to define a non-parametric estimate 
of the cluster probabilities. We decided however to use them to define a 
Gaussian mixture model in which each cluster corresponds to one of the sets.\footnote{Quick shift is 
a \emph{medoid} method, and as such the modes it finds are bound to be grid points. This means
that the resolution of the grid impacts the definition of the cluster centers. In practice,
this is easily solved by quickly optimizing the positions of the points $\mathbf{z}_k$ using
mean shift.}
Gaussian mixture models\cite{press07book,reyn09book} are a particularly convenient approach to 
parametrically fit the probability distribution as a sum of $n$ multivariate Gaussians,
\begin{equation}
\hat{P}(\mathbf{x})=\sum_{k=1}^{n}p_{k}G\left(\mathbf{x}\middle|\boldsymbol{\mu}_{k},\boldsymbol{\Sigma}_{k}\right),
\label{eq:gmm}
\end{equation}
where $p_{k}$ is the fraction of density belonging to the $k$-th Gaussian 
(the relative weight of the cluster), $\boldsymbol{\Sigma}_{k}$ is its covariance 
matrix and $\boldsymbol{\mu}_{k}$ the mean:
\begin{equation}
G\left(\mathbf{x}\middle|\boldsymbol{\mu},\boldsymbol{\Sigma}\right) = \frac{1}{\sqrt{(2\pi)^D\det \boldsymbol{\Sigma}}}
e^{ {-\left(\mathbf{x}-\boldsymbol\mu\right)^T\boldsymbol{\Sigma}^{-1} \left(\mathbf{x}-\boldsymbol\mu\right)}/{2} }.
\end{equation}

Fitting a parametric model such as~\eqref{eq:gmm} to the probability density lends itself
quite naturally to a probabilistic interpretation in which each Gaussian function corresponds
to a cluster, so that one can evaluate the probability that a point $\mathbf{x}$ belongs
to the $k$-th cluster as 
\begin{equation}
\hat{P}_k(\mathbf{x})=p_k G\left(\mathbf{x}\middle|\boldsymbol{\mu}_{k},\boldsymbol{\Sigma}_{k}\right) / \hat{P}(\mathbf{x}).
\label{eq:gmm-pk}
\end{equation}
In the context of the present work, if we associate one of the Gaussian clusters of the density
of sample points with the configuration pattern that we are trying to recognize, this
cluster probability provides a very natural, fuzzy definition of the region of parameter space
associated with the recurring pattern. $\hat{P}_k(\mathbf{x})$ varies smoothly between zero
(configurations that clearly differ from the tagged cluster) to one (configurations that are 
undoubtedly to be assigned to the cluster), with intermediate values representing the 
``gray zone'' at the margins of the cluster. 
\blue{For instance, as we shall see in Section~\ref{sec:results}, if one can assign the 
$k$-th mode of $P(\nu,\mu,r)$ to hydrogen-bonded configurations, $\hat{P}_k(\nu,\mu,r)$
would give a data-driven hydrogen-bond count function, that would have a smooth transition
between a value of 0 for non-hydrogen-bond configurations to 1 for configurations that clearly
belong to the HB region of parameter space.}

It is customary to train Gaussian mixture models by fitting the parameters $p_k$, 
$\boldsymbol{\mu}_{k}$, $\boldsymbol{\Sigma}_{k}$  so as to maximize the log-likelihood
of the model given the underlying density $P(\mathbf{y}_i)$,  
\begin{equation}
\mathcal{L} = \sum_i P(\mathbf{y}_i) \ln \hat{P}(\mathbf{y}_i).
\end{equation}
\blue{This idea has been exploited before in the context of atomistic simulations, see 
e.g. Refs.~\cite{mara+09jpcb,trib+10pnas,pisa+14jctc}.}
We found however this procedure to be unreliable for our purposes: the optimized parameters 
depend dramatically on the number of clusters included in the model and on the initial values 
assigned to the parameters. Furthermore, if the modes of the probability distribution are not well described
by Gaussians, it is often the case that more than one Gaussian is needed to describe each cluster,  
requiring further fine-tuning of the model. 

To avoid this ambiguity, we did not obtain the Gaussian parameters by optimizing the log-likelihood of 
the model, but rather took one Gaussian for each of the clusters identified by quick shift, setting its
mean at the maximum of the probability distribution and its covariance based on the distribution of the 
cluster members $\mathcal{Z}_k$ (Figure~\ref{fig:pamm}(e)):
\begin{equation}
\begin{split}
\boldsymbol{\mu}_k=\mathbf{z}_k, \quad
p_k=\sum_{\mathbf{y}\in\mathcal{Z}_k} P(\mathbf{y})/\sum_{\mathbf{y}\in\mathcal{Y}} P(\mathbf{y}), \\
\boldsymbol{\Sigma}_k={\sum_{\mathbf{y}\in\mathcal{Z}_k} P(\mathbf{y}) \left(\mathbf{y}-\mathbf{z}_k\right)\left(\mathbf{y}-\mathbf{z}_k\right)^T}/
{ \sum_{\mathbf{y}\in\mathcal{Z}_k} P(\mathbf{y}) }.
\end{split}
\end{equation}
This choice combines the simplicity of a Gaussian mixture model, the fuzzy, smooth nature
of the cluster probabilities~\eqref{eq:gmm-pk}, and the robust, deterministic 
partitioning of the probability density obtained with quick shift. 

\subsection*{Analysis of the simulation}

A critical analysis of the outcome of the quick shift partitioning of the probability 
density makes it possible to associate each cluster with a structural pattern,
as described by the $D$ dimensional vector introduced in the first phase of our
procedure. In many cases -- such as the example of the hydrogen bond that we will
discuss further below -- one cluster stands out clearly as the distinct structural
feature one is interested in, and one can focus further analysis on a single mode 
using the associated conditional probability function~\eqref{eq:gmm-pk}. 
For simplicity we will consider the selected cluster to be the one labeled by $k=1$.  

The most direct application of the definition embodied by $\hat{P}_1(\mathbf{x})$ is to use
it to test whether tuples of atoms $\mathbf{R}_{ijk\ldots}=(\mathbf{r}_i,\mathbf{r}_j,\mathbf{r}_k,\ldots)$ 
match the definition of the structural pattern by computing
\begin{equation}
s^{(D)}_{ijk\ldots}=\hat{P}_1(\mathbf{x}(\mathbf{R}_{ijk\ldots})). \label{eq:sijk}
\end{equation}
If the components of $\mathbf{x}$ are continuous functions of the atomic coordinates,
$\hat{P}_1$ is a smooth continuous function too, that takes a value close to 1 whenever
a group of atoms matches the target pattern (Figure~\ref{fig:pamm}(f)). This makes our definition of a pattern 
recognition function well-suited for use as a collective variable in accelerated sampling 
methods,\cite{laio-parr02pnas} possibly in conjunction with other machine learning
techniques to characterize the overall connectivity induced by the selected molecular
pattern~\cite{piet-andr11prl}, \blue{or similar fingerprint metrics that are guaranteed
to distinguish dissimilar structures~\cite{sade+13jcp}}.
The PAMM variables corresponding to different structural descriptor can also be analyzed to yield
a coarse-grained, low-dimensional map~\cite{ferg+10pnas,trib+12pnas,rohr+arpc03}.
If necessary, one can artificially ``soften'' the transition between clusters, by dividing
\emph{all} the covariance matrices $\boldsymbol{\Sigma}_k$ in the Gaussian model by a scaling 
factor $\alpha$.

 Since the $s^{(D)}_{ijk\ldots}$'s effectively count the instances of the structural pattern
that are present at any given time in the trajectory, one can also combine several 
of these indicators together to count the number of patterns that involve a 
tagged atom $i$, or pair of atoms $\left(i,j\right)$:
\begin{equation}
s^{(1)}_{i}=\sum_{j,k,\ldots} s^{(D)}_{ijk\ldots},\quad
s^{(2)}_{ij}=\sum_{k,\ldots} s^{(D)}_{ijk\ldots}, \quad\text{etc.}
\end{equation}
Depending on the application being considered, $s^{(1)}_{i}$ can be 
taken to represent the total coordination of the atom $i$, $s^{(2)}_{ij}$ the overall 
bonding between atoms $i$ and an atom $j$, and so on.
\blue{For instance, by summing over all the possible acceptor atoms O$'$ and hydrogen
atoms H, one can get a smooth count of the total number of HBs donated by a selected
oxygen O.}

\section{The hydrogen bond, revisited\label{sec:results}}

The PAMM framework we have introduced in the previous section is very abstract,
and can be applied to any situation in which one wishes to recognize recurring
motifs in an atomistic simulation. To demonstrate its application in a practical
case, we chose to focus on recognizing the hydrogen bond (HB) in a number of different
contexts. The term ``hydrogen bond'' refers to a highly directional three-centers 
interaction between two polar atoms and a hydrogen.\cite{desi11ac,arun+11PureApplChem} 
The hydrogen atom H is covalently 
bound to one of the polar atoms, which is designated as the donor D, and points
towards the second polar atom which is designated as the HB acceptor A. 
Despite the apparent simplicity of the concept, it is not easy to develop
an universal definition of the HB, mostly because this entity has been used 
in many different contexts. The term has been associated to near-covalent interactions
with an energy in excess of 30 kcal/mol, as well as to exceedingly weak ones with an energy
of less than one kcal/mol. Typically hydrogen bonds are understood to have a 
predominantly electrostatic nature, with strongly electronegative donors and 
acceptors such as F, O or N. However, the observation of recurring \HB{C}{O} units
in the secondary structure of polypeptides have also been interpreted in terms 
of weak HBs, that have been suggested to play a significant role in 
stabilizing proteins.\cite{dere+95JMolBiol}

Adding to the complexity of the broad energy scale covered by 
HBs is the fact that in most situations of interest thermal fluctuations
and the environment modulate their stability, and that they are formed and 
destroyed on a relatively short time scale. One sees the difficulty in
giving a clear-cut definition of a chemical entity which exhibits 
such a variability. The most generally applicable definitions rely on
performing an electronic structure calculation, and on decomposing the 
energy of the systems in a sum of terms that can be interpreted as 
the binding energy of putative HBs.\cite{reed+1988ChemRev,pend+2006JChemPhys,azarhead12JChemPhys,kuhn-khal14jacs}
Definitions that are based solely on structural information are much 
more practical, in that they do not require a supporting electronic 
structure calculation and can be applied to experimental structural
data or to atomistic simulations based on empirical forcefields.
The downside is that these structural definitions invariably contain
a degree of arbitrariness, as they are based on the heuristic 
introduction of ranges of structural parameters that are deemed to 
represent a hydrogen bond in a given context.\cite{tayl-kenn84AccChemRes,mats2007JChemPhys} 
Kumar \emph{et al.} carried out a systematic comparison of many
of these structural definitions in the case of liquid water,\cite{kuma+07jcp} 
and recognized that the best way to assess whether a given definition 
makes physical sense is to compare the probability distribution of the 
structural parameters with the range of values associated with the 
hydrogen bond.

The PAMM algorithm we have introduced in Section~\ref{sec:methods} 
makes this probabilistic analysis the very basis of the construction, 
automatically determining the range of structural parameters that
corresponds to one of the modes of the distribution. 
The data itself informs the definition of a range of parameters that
identify unambiguously hydrogen-bonded configurations, and naturally
describes smoothly the transition between this region and configurations
that are clearly not hydrogen bonded. Even though this definition is 
by construction system-specific, the protocol to obtain it is univocal
and unbiased, as it does not rely on choosing manually threshold values 
for the structural parameters.

The first steps in the application of PAMM are the identification of groups of 
atoms that should be tested for recurring patterns and the choice of 
structural parameters that describe the arrangement of atoms within each group.
In the case of the HB, these choices are fairly obvious. One should select
an atomic species that should be considered as the putative HB donor D,
one that should be considered as the acceptor A and (a subset of) the hydrogen
atoms that complete the HB triplet. The geometry of each of these groups
is completely determined by the three distances $d(\text{A-D})$, $d(\text{A-H})$
and $d(\text{D-H})$. To simplify comparison with other definitions, and 
to highlight the symmetries inherent in the problem, we decided to use
combinations of these distances, namely the proton-transfer coordinate
$\nu=d(\text{D-H})-d(\text{A-H})$, the symmetric stretch coordinate
$\mu=d(\text{D-H})+d(\text{A-H})$ and the acceptor-donor distance
$r=d(\text{A-D})$ as the group descriptors. 
We computed these $(\nu,\mu,r)$ triplets for each D-H-A group present in 
each snapshot extracted  from the simulations, thereby obtaining
the training data set $\mathcal{X}$ that we used to run PAMM.
In building the probability distribution, each point was weighed
by a factor $\left[r(\nu+\mu)(\mu-\nu)\right]^{-1}$, that accounts
for the trivial phase space volume so that a uniform distribution of 
atoms would yield a constant probability density in $(\nu,\mu,r)$.

\subsection*{Alanine dipeptide}

Let us begin our analysis by considering the case of an empirical forcefield 
model of alanine dipeptide (N-acetylalanine-N'-methylamide) -- one of the simplest examples of peptide 
bonding, displaying many of the essential features that are present in proteins.
In our case, it is an ideal test case, as it allows us to demonstrate the 
functioning of PAMM for different kinds of hydrogen bonds. We will 
consider HBs donated by water molecules to the carbonyl of alanine O$_\text{C}$, 
HBs donated by the peptide nitrogen to the oxygens in water O$_\text{w}$, and investigate
the significance of a more exotic, weak HB donated by the peptide C$_\alpha$ to the O$_\text{w}$ atoms.

We used the CHARMM27 forcefield\cite{mack+98JPhysChemB} to describe interactions within the polypeptide and a 
TIP3P model for the water molecules,\cite{jorg83JChemPhys} with flexible bonds modelled as harmonic stretches,
as implemented in LAMMPS.\cite{plim95jcp}
We equilibrated a supercell containing 128 water molecules in the \emph{NpT} ensemble,
and ran subsequently 600~ns of \emph{NVT} molecular dynamics using a Langevin thermostat with a
time constant of 10~ps.\cite{schn-stol78prb}  
The configurations were saved every 1~ps.
An example input file is provided in the SM\cite{si}.

\begin{figure}[htbp]
        \caption{\label{fig:ala-OHO}(Upper panel) Distribution of $(\nu,\mu,r)$ configurations for \HB{O$_\text{w}$}{O$_\text{C}$}
        in a simulation of alanine dipeptide in water. Size and opacity of points correspond to the KDE of
        $P(\mathbf{y})$, and colors indicate the cluster each grid point has been assigned to.
        (Lower panel) Free energy (kcal/mol) computed from the distribution of number of accepted hydrogen bonds $s_\text{A}$
        for the oxygen atom in the carbonyl group in solvated alanine dipeptide. The histogram was smoothed
        with a triangular kernel of width 0.025. We also report the integrated probabilities for 
        for having $s_\text{A}<0.5$, $0.5<s_\text{A}<1.5$ and so on. 
        The average number of accepted HBs is $\left<s_\text{A}\right>=2.1$ and the standard deviation is 0.6. 
}

        \centering
        \includegraphics[width=\onecol]{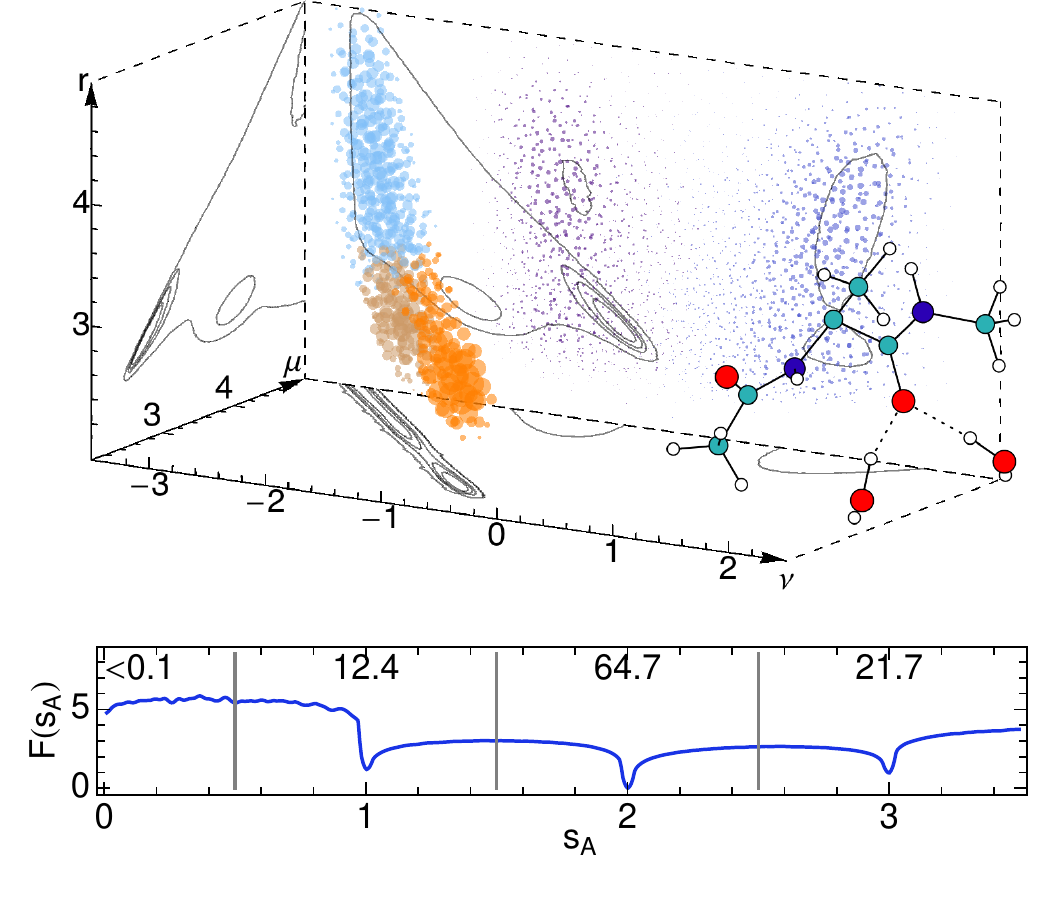}
\end{figure}

The first kind of HB we considered is \HB{O$_\text{w}$}{O$_\text{C}$}. To accelerate
the analysis we only included configurations with $\mu<5$\AA{}. Figure~\ref{fig:ala-OHO}
shows the distribution of the values of  $(\nu,\mu,r)$, colored according to the partitioning
of the density obtained by running PAMM on the data set. Several clusters are recognized, 
which means that besides HBs there are other recurring patterns that can be distinguished 
by this analysis. One of these clusters -- represented with an orange hue -- can be seen by
direct inspection of configurations (or by comparison with other structural definitions)
to correspond clearly to hydrogen-bonded configurations. As discussed in Section~\ref{sec:methods}
we used the Gaussian-mixture model built based on the clustering to define the 
degree of confidence $s_\text{DHA}$ by which we classify a certain configuration of a D donor, H hydrogen
and A acceptor as a hydrogen bond, and then introduce a count of the total HBs that involve a given acceptor 
oxygen $s_\text{A}=\sum_{\text{D},\text{H}} s_\text{DHA}$. The free energy built from the histogram of $s_\text{A}$
is represented in the lower panel of Figure~\ref{fig:ala-OHO}.
The free energy is strongly peaked at  integer values of $s_\text{A}$, because the transition 
between 0 and 1 is very sharp when a hydrogen bond is formed or broken. This plot shows clearly 
that most of the time the carbonyl is involved in receiving two hydrogen bonds, but there is also a 
fairly large probability of accepting one or three bonds. It would be interesting to compare these 
results with first-principles simulations of solvated alanine dipeptide, to verify whether the possibility
of forming over and under-coordinated configurations is a consequence of the simplified modelling of the 
interactions between water molecules and the carbonyl.

\begin{figure}[htbp]
        \caption{\label{fig:ala-NHO}(Upper panel) Distribution of $(\nu,\mu,r)$ configurations for \HB{N}{O$_\text{w}$}
        in a simulation of alanine dipeptide in water. Size and opacity of points correspond to the KDE of  
        $P(\mathbf{y})$, and colors indicate the cluster each grid point has been assigned to.
        (Lower panel) Free energy (kcal/mol)  computed from the distribution of number of donated hydrogen bonds $s_\text{D}$
        for the amide nitrogen atom in solvated alanine dipeptide. See Fig.~\ref{fig:ala-OHO} for details.
        The average number of donated HBs is $\left<s_\text{D}\right>=1.3$ and the standard deviation is 0.5. 
}
        \centering
        \includegraphics[width=\onecol]{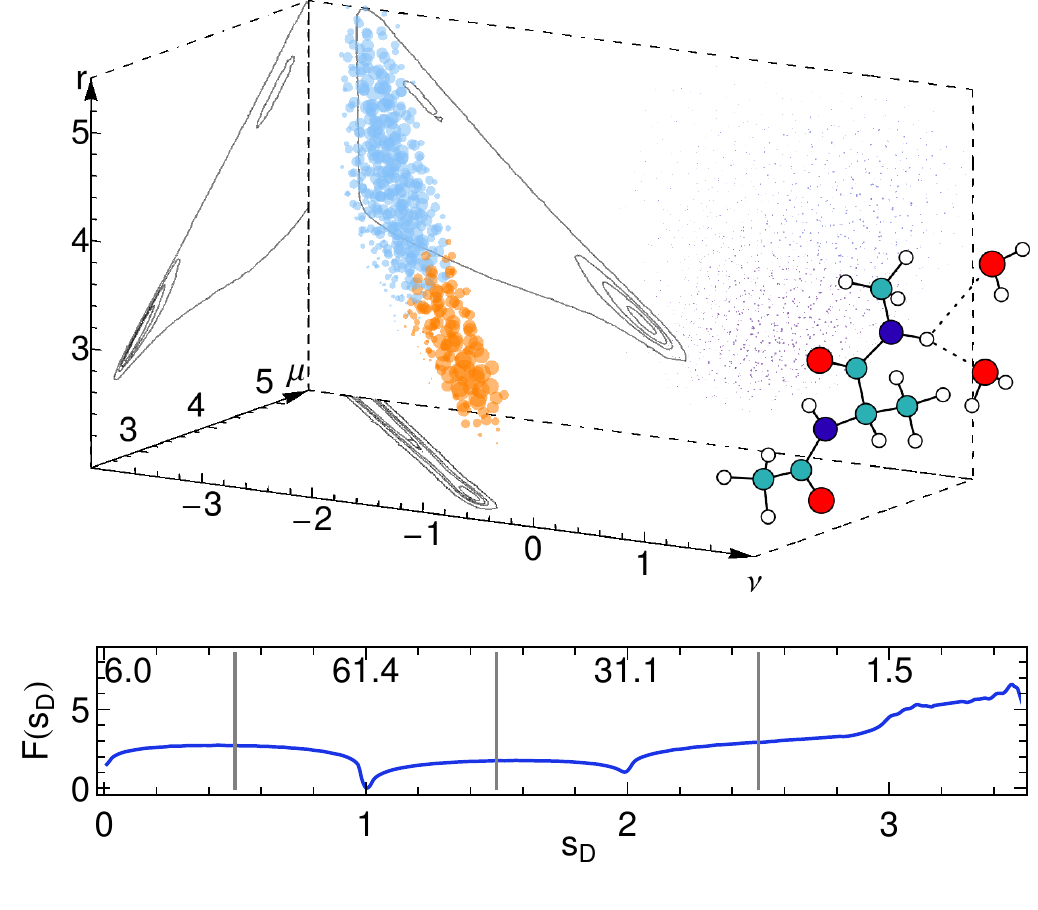}
\end{figure}

We then moved on to look into the hydrogen bond donated by the amide group
\HB{N}{O$_\text{w}$}. Since the chemical identity of atoms is fixed in an 
empirical force field calculation, we specifically restricted the search to include only
the amide H atom and oxygen atoms from the water molecules.
We used a cut off of  $\mu<5.5$\AA{} to disregard configurations that are clearly 
irrelevant to the HB search. We report the 
distribution of configurations and the PAMM clustering in Figure~\ref{fig:ala-NHO}.
Perhaps unsurprisingly, the distribution of $(\nu,\mu,r)$ associated with this set
of atoms differs considerably from that in Figure~\ref{fig:ala-OHO} -- 
this is a somewhat weaker bond, which results in a less structured $P(\nu,\mu,r)$. 
Still, one can recognize a cluster that is clearly associated with HB configurations,
that we can use to define a bond counting order parameter, that in turns can be used to compute 
the total number of hydrogen bonds donated by the N atom, 
$s_\text{D}=\sum_{\text{A},\text{H}} s_\text{DHA}$.
While the most likely value of $s_\text{D}$ is one, there is a high probability of observing 
a N--H group donatinhg two HBs. Given the geometry of the amide group, this
actually means that based on our unbiased, self-consistent definition, HBs donated by the amide
group as described by the empirical forcefield we used have a large probability of being 
bifurcated, binding simultaneously to two different water molecules. 

\begin{figure}[htbp]
        \caption{\label{fig:ala-CHO}(Upper panel) Distribution of $(\nu,\mu,r)$ configurations for \HB{C$_\alpha$}{O$_\text{w}$}
        in a simulation of alanine dipeptide in water. Size and opacity of points correspond to the KDE of the 
        $P(\mathbf{y})$, and colors indicate the cluster each grid point has been assigned to.
        (Lower panel) Free energy computed from the distribution of number of donated hydrogen bonds $s_\text{D}$
        for the amide carbon atom in solvated alanine dipeptide. See Fig.~\ref{fig:ala-OHO} for details.
        The average number of donated HBs is $\left<s_\text{D}\right>=5.2$ and the standard deviation is 1.1. 
}
        \centering
        \includegraphics[width=\onecol]{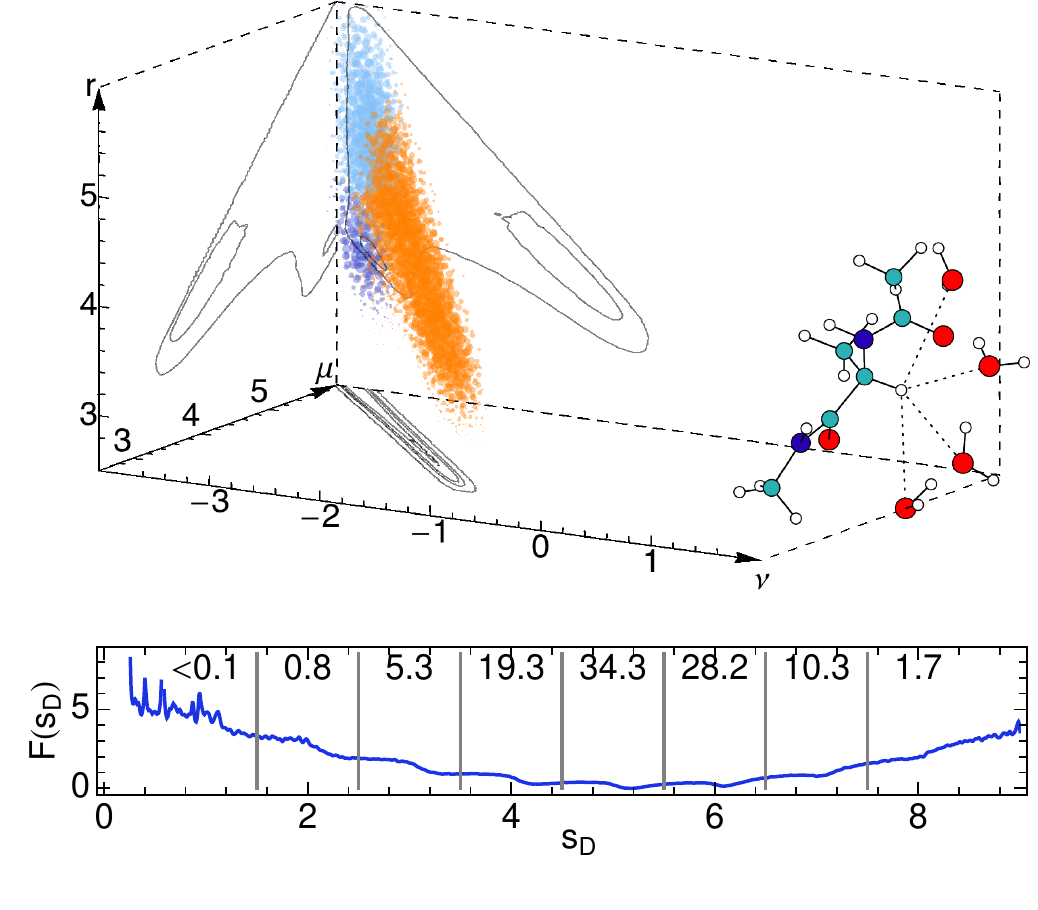}
\end{figure}

Finally, to verify how the PAMM algorithm behaves when applied to a selection of atoms 
that does not exemplify a typical hydrogen bond, we considered atom triplets that would  
correspond to a C$_\alpha$--H group donating a HB to water. Figure~\ref{fig:ala-CHO} shows the 
partitioning of the probability density for this choice of atoms, which has 
some features that are reminiscent of those seen for conventional HBs, albeit
with a much longer $d(\text{A--H})$. A more careful inspection of configurations that 
belong to the lobe of the probability density with the lowest $\mu$, however, 
shows that these can hardly be described as HBs: in many cases the hydrogen atoms of water
molecules are oriented \emph{towards} the C$_\alpha$--H group, and the distribution
of $s_D$ shows very little structure. 
This example demonstrates that the presence of a recurring structural motif with a 
signature in terms of the probability distribution in configuration space does not 
necessarily imply that the atoms that compose the motif are involved in some sort
of chemical bonding. Here, the non-uniform structure of oxygen atoms in the vicinity of the 
 C$_\alpha$--H group is probably an indirect consequence of the hydrogen-bond interaction of 
water molecules with nearby carbonyl groups, and of the stiffness of the backbone
of the dipeptide.

\subsection*{Water, classical and quantum}

\begin{figure}[htbp]
        \centering
        \includegraphics[width=\onecol]{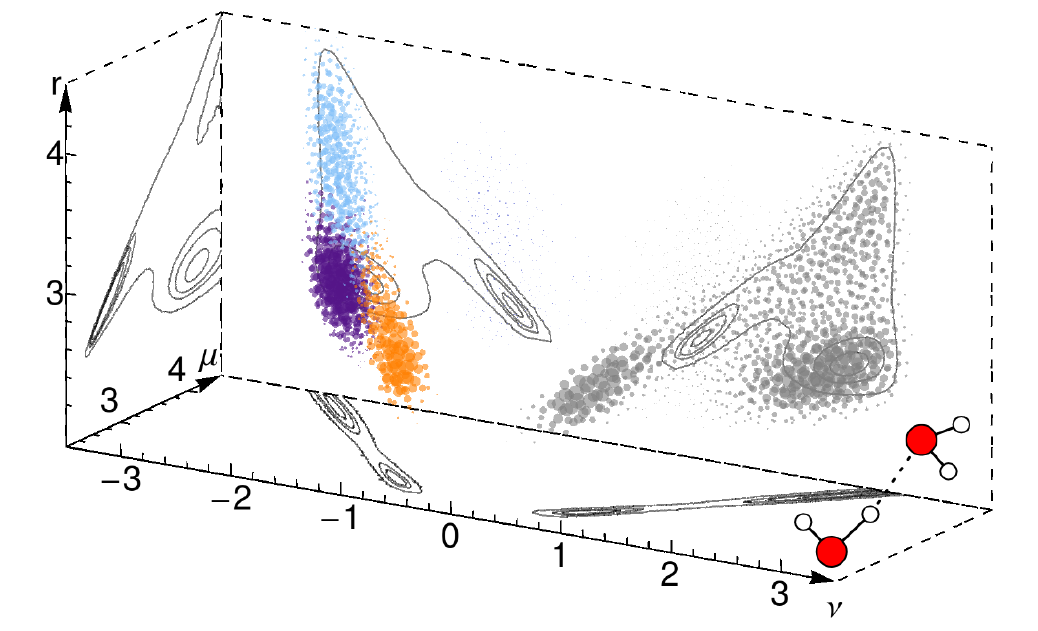}
        \caption{\label{fig:h2o-3d} Distribution of $(\nu,\mu,r)$ configurations for \HB{O}{O$'$}
        in a simulation of neat TIP4P water. Size and opacity of points correspond to the KDE of 
        $P(\mathbf{y})$, and colors indicate the cluster each grid point has been assigned to.
        Clusters with $\nu>0$ have not been colored, but have been correctly identified by PAMM.
        }
\end{figure}

Simulations of alanine dipeptide contain different kinds of hydrogen bonds, and allowed us to 
demonstrate the adaptive nature of PAMM to derive a different, data-driven definition of the
range of structural parameters that can be associated with a HB for each set of constituent
atoms. In a simulation of neat water, instead, there is only one type of \HB{O}{O$'$}, the slight 
complication being that each oxygen atom can simultaneously act as a donor and an acceptor of
hydrogen bonds. 

\paragraph{TIP4P water.}
We began by analyzing a simulation of a flexible TIP4P/2005f\cite{gonz+11JChemPhys} model.
A box containing 128 water molecules was first equilibrated for 2~ns at constant pressure (1 atm),  
constant temperature (298 K) \emph{NpT} dynamics. A subsequent 500~ns \emph{NVT} run was performed 
using a Langevin thermostat with relaxation time $\tau$=5~ps. The configurations were saved every 1~ps.
In the spirit of a fully automated analysis of 
the trajectory, we did not exploit knowledge of the chemical identity of water molecules, which
is fixed in a simulation with a non dissociable model. The distribution of $(\nu,\mu,r)$ 
shows clearly the dual role played by the O atoms (see Figure~\ref{fig:h2o-3d}), which is
apparent in the symmetry of the probability density across the $\nu=0$ plane. Both the 
cluster highlighted in orange and its mirror image correspond to legitimate HB configurations, but 
only the former corresponds to structures in which the first oxygen is acting as the donor and the 
second as the acceptor. 

\begin{figure}[htbp]
        \centering 
        \includegraphics[width=\onecol]{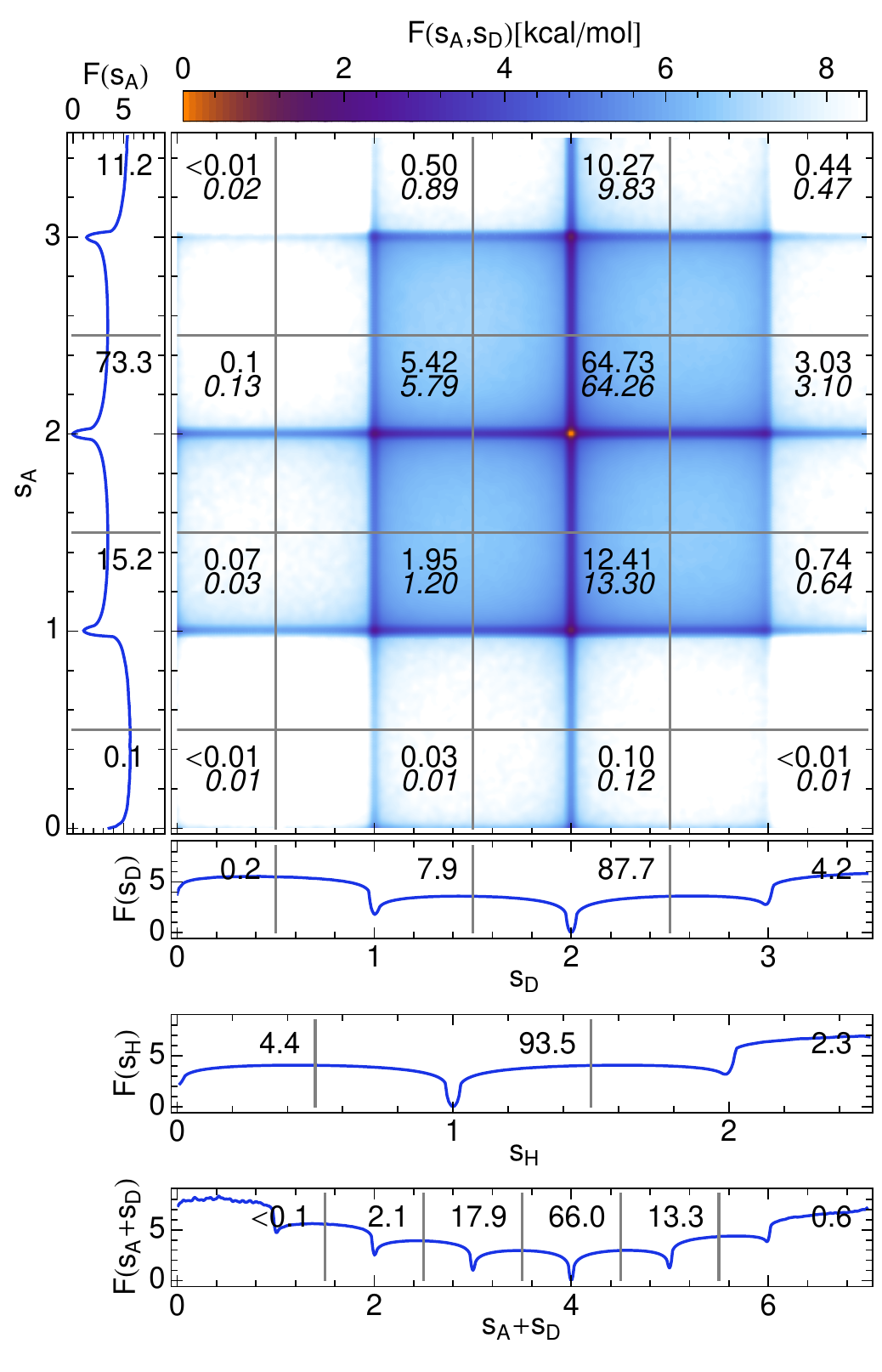}
        \caption{\label{fig:h2o-tip4p-cls} Hydrogen-bond counts statistics for a classical 
simulation of TIP4P water at room temperature. All the probability distributions have been
smoothed with a triangular kernel of width 0.025, and are represented in terms of the associated
free energies $F=-k_BT\ln P$, that are expressed in kcal/mol throughout. We also report integrated
probabilities (in percent) to have a configuration in the vicinity of the different integer 
numbers of HBs. Below the values of the joint probabilities of $s_\text{A}$ and $s_\text{D}$ 
the product of marginal probabilities are indicated, in italics.}
\end{figure}

Once $s_\text{DHA}$ has been defined based on the analysis of the simulation data, it can be used to characterize
in great detail how a given model of water describes hydrogen bonding. 
Figure~\ref{fig:h2o-tip4p-cls} summarizes some of the information that can be obtained from this analysis.
One can compute free energies for the number of hydrogen bonds donated ($s_\text{D}$) or accepted
($s_\text{A}$) by each oxygen atom, as well as for the total ($s_\text{A}+s_\text{D}$) and for the
number of HBs that are formed by each hydrogen. 
A large fraction of TIP4P water molecules are tetracoordinated, with nearly 65\%{} of oxygen 
atoms receiving and donating two HBs. There is a small but significant asymmetry between the 
distribution of $s_\text{D}$ and that of $s_\text{A}$, the former being more strongly peaked at 
$s_\text{D}=2$, while there is somewhat more flexibility in the count of accepted bonds. 
Given the rigid constraints on the covalent O--H bond, any oxygen in the simulation
can donate two bonds at most, except for the case of bifurcated HBs where a single O--H moiety is
involved with bonds to two different O$'$ atoms. The distribution of $s_\text{H}$ shows that there is 
just about 2\%{} probability of observing such bifurcated bonds. 
 More detailed information on the topology of the HB network can be obtained by observing the joint 
probability distribution of $s_\text{D}$ and $s_\text{A}$. While the order of magnitude of the probability of 
each joint configuration is determined by the product of  $s_\text{D}$ and $s_\text{A}$, there are significant 
deviations that are indicative of the correlations between defects in the network. For instance,
the probability of having a ``linear'' water that donates and accepts a single HB is almost twice the value
that would be expected based on the product of the marginal distributions. 
Note that this analysis focuses on the connectivity of the network rather than on the geometry
of the environment of each water molecule. An interesting way to extend this analysis could consider
the correlation between the HB counts and the degree of tetraedrality, with the electronic structure, or 
with quantities that can be directly related to experimental observables.

\begin{figure}[htbp]
        \centering
        \includegraphics[width=\onecol]{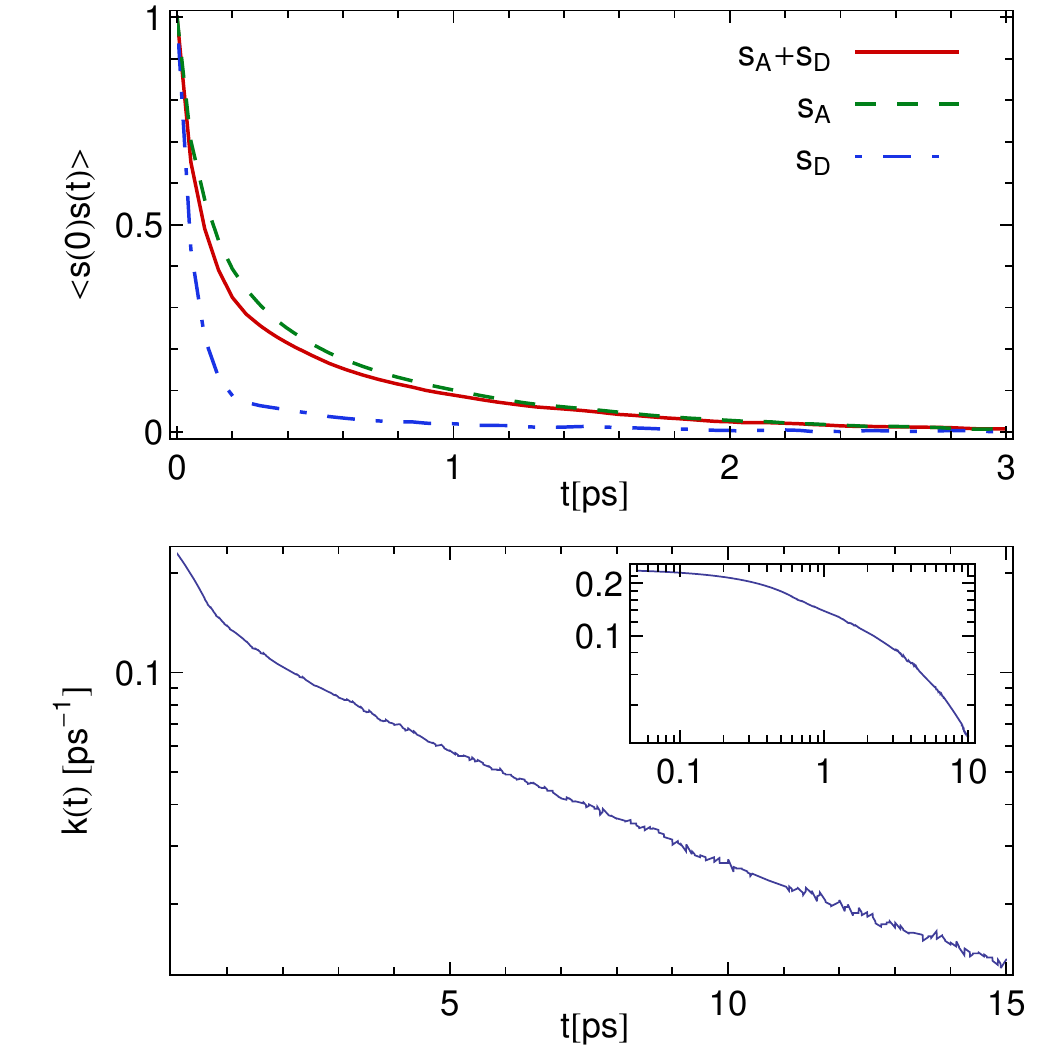}
\caption{\label{fig:h2o-nhb-acfs-tip4p} (Upper panel) Time correlation functions of the hydrogen-bond
counts for a tagged oxygen in a simulation of TIP4P/2005f water. (Lower panel) Rate function for 
the hydrogen-bond formation/break-up, computed as the derivative of the correlation function 
$\left<s_\text{DA}(t)s_\text{DA}(0)\right>$ }
\end{figure}

\paragraph{Dynamics of the HB network.}
The asymmetry between $s_\text{A}$ and $s_\text{D}$ is also apparent when one considers
the dynamical behavior of the two quantities. The upper panel of Figure~\ref{fig:h2o-nhb-acfs-tip4p} 
shows the correlation functions for the counts of acceptor and donor HBs for a 
tagged oxygen, as well as the total. 
They were computed analyzing several short \emph{NVE} simulations started from 
independently equilibrated configurations. The correlation time of $s_\text{D}$ is very short, 
because configurations where a water molecule donates less or more than two HBs are very short-lived.
Configurations that are distorted from the point of view of accepted HBs are less unstable,
and therefore the correlation function of $s_\text{A}$ decays more slowly. 
The correlation function of the total is dominated by the slow decay of $s_\text{A}$,
and is compatible with the results reported in Ref.\cite{kuma+07jcp} for traditional
structural definitions of the hydrogen bond.

The hydrogen-bond count functions we have used this far do not consider the identity of 
individual bonds, so that a quick fluctuation that momentarily breaks a HB that is 
immediately re-formed is indistinguishable from a fluctuations that breaks a HB and leads immediately to the
formation of a new HB with a \emph{different} acceptor oxygen. A correlation function 
that is sensitive to the identity of the HB triplet, which is more easily interpreted
in terms of physical observables,\cite{luza-chan96nature} can be readily computed by 
considering all the $(\text{D},\text{A})$ pairs, computing for each pair
$s_\text{AD}=\sum_\text{H} s_\text{DHA}$. One can then compute the 
rate function as the time derivative of the autocorrelation 
of $s_\text{AD}$, computed for each pair separately:
\begin{equation}
k\left(t\right)=-\frac{1}{n_\text{A}n_\text{D}}\frac{\partial}{\partial t} \sum_\text{A,D}\left<s_\text{AD}(t)s_\text{AD}(0)\right>.
\label{eq:k-hb}
\end{equation}
The decay takes place on a similar time scale than observed in Ref.~\cite{luza-chan96nature}, 
and exhibits similar features, including the presence of multiple time scales in the decay
of the rate function. 

\begin{figure}[htbp]
        \centering
        \includegraphics[width=\onecol]{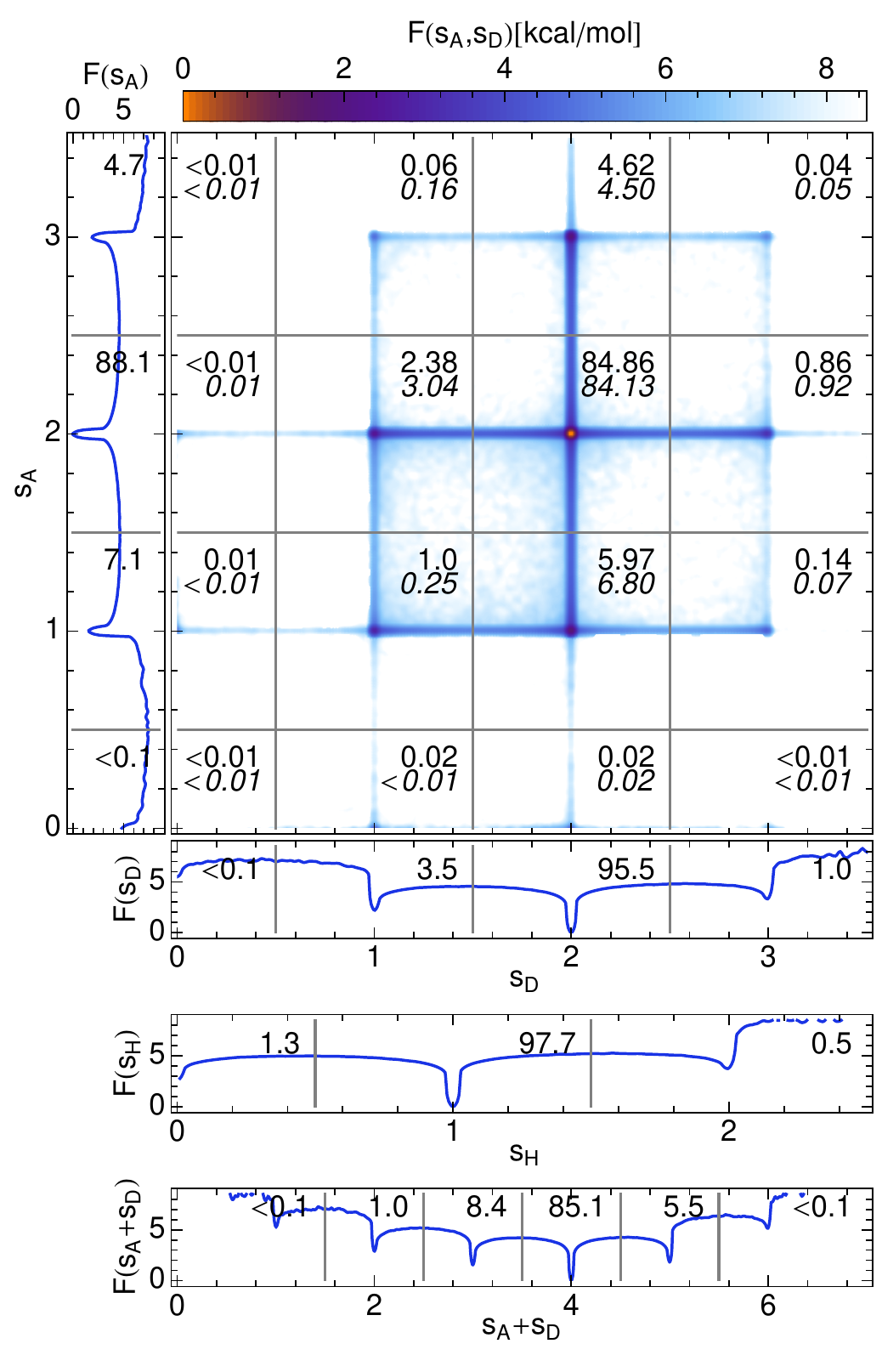}
        \caption{\label{fig:h2o-blyp-cls} Hydrogen-bond counts statistics for a classical 
simulation of BLYP water at room temperature. See the caption of Figure~\ref{fig:h2o-tip4p-cls}
for a detailed explanation of the plots.}
\end{figure}

\blue{
\paragraph{Hydrogen-bonding defects in ice Ih}

The consistency of the results obtained with PAMM descriptors of the hydrogen bond and those obtained 
with more conventional descriptors is reassuring. It is however useful to verify the behavior of 
indicators such as $s_\text{A}$ and $s_\text{D}$ in a more ordered environment such as the 
tetrahedral hydrogen-bond network of ice Ih, in which one should be able to identify clearly
coordination defects. 
}

\begin{figure}[tbhp]
    {\centering \includegraphics[width=0.75\columnwidth]{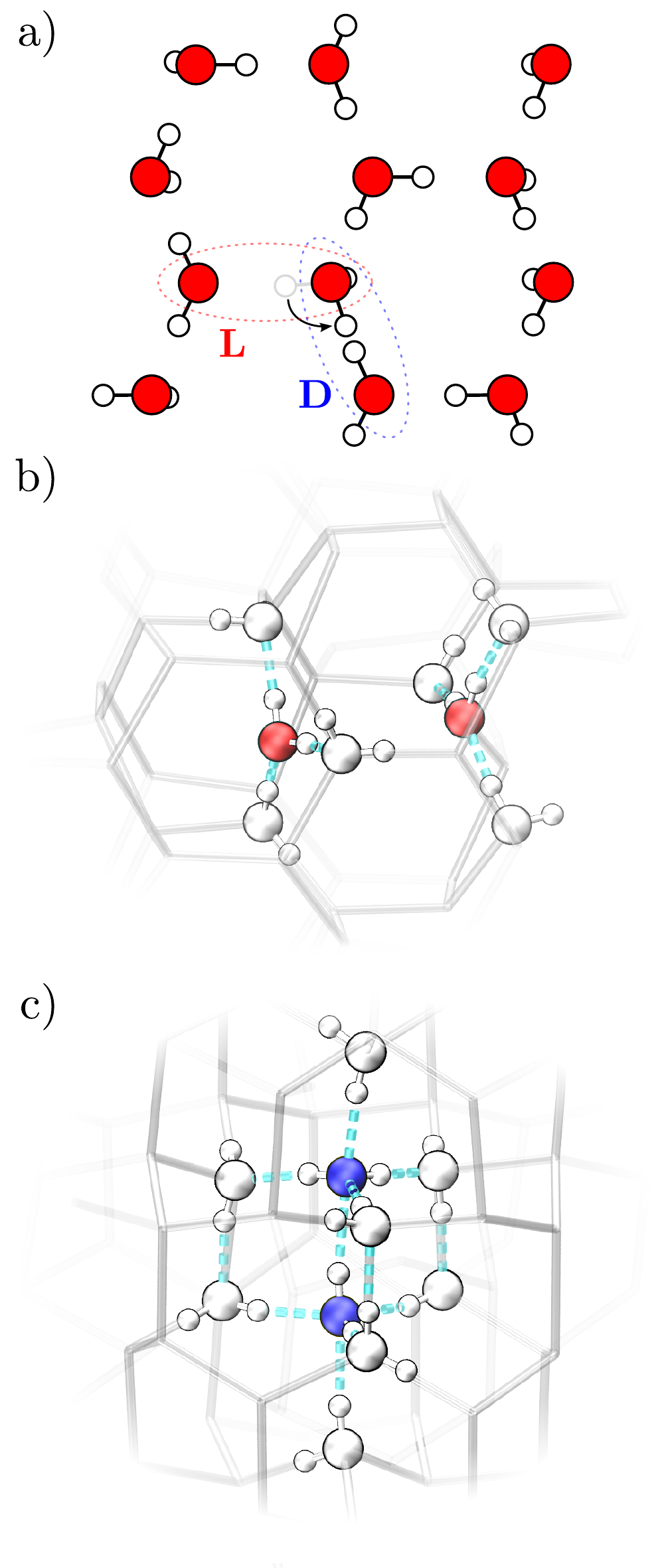} }  
    \caption{\label{fig:bjerrum} \blue { a) Schematic representation of how a pair of Bjerrum defects 
    are generated by flipping the orientation of a water molecule in a lattice that satisfies the 
    ice rules. By flipping other molecules, the L and D defects can be separated. b) and c) Configurations of 
    equilibrated L  and D defects in ice Ih. Oxygen atoms with $s_\text{A}=1$ are colored in red, atoms
    with $s_\text{A}=3$ are colored in blue, while atoms with
    $s_\text{A}=s_\text{D}=2$ are in white, or hidden for clarity. }}
\end{figure}

\blue{
To this aim, we have performed a PAMM analysis of a simulation of ice Ih, using the same 
flexible TIP4P model discussed above, and a proton-disordered unit cell with 768 molecules~\cite{hayw-reim97jcp}. 
We have then created a pair of Bjerrum coordination defects~\cite{bjer52science,deko+2006prl}, and separated them by the maximum
distance allowed by the simulation cell, by repeatedly flipping water molecules in the lattice (Figure~\ref{fig:bjerrum}(a)).
We have then equilibrated the simulation for a few tens of ps, collected some snapshots of the 
configurations and evaluated $s_\text{A}$ and $s_\text{D}$ for each oxygen. 
The vast majority of the O atoms have $s_\text{A}=s_\text{D}=2$, as one would expect in a 
perfect tetrahedral arrangement consistent with the ice rules. We could however identify clearly
a pair of oxygen atoms with $s_\text{A}=1$ (a L defect), and a pair with $s_\text{A}=3$ (a relaxed D defect).
Snapshots of these defective environments are represented in Figure~\ref{fig:bjerrum}.
}

\paragraph{Classical \emph{ab initio} water.}
We then moved on to perform our analysis on a first-principles simulation of liquid water.
The trajectory is the classical simulation from Ref.~\cite{ceri-mano12prl}, which was performed
using the CP2K software package,\cite{vand-krac05cpc,*goed+96prb} 
with a BLYP exchange-correlation functional\cite{beck88pra,*lee+88prb} and a DZVP basis set.
The simulation box contained 64 water molecules at the experimental density, and 100~ps of
\emph{NVT} dynamics were performed, with the first 5~ps discarded for equilibration. 
A PAMM analysis of the simulation yielded very similar clusters to those obtained from
TIP4P water. The analysis of HB counts, in Figure~\ref{fig:h2o-blyp-cls}, shows that BLYP 
water has a very regular structure, with a higher count of tetracoordinated oxygen atoms,
and a very low count of defective structures. This is consistent with the well-known 
observation that generalized-gradient approximation models of water are overstructured 
compared to experiment and to empirical water models. Note that correlations in the HB
network are stronger in this case than for TIP4P water, with one-donor/one-acceptor oxygen
atoms being four times more likely than one would expect given the separate probabilities
of $s_\text{A}\approx 1$ and $s_\text{D}\approx 1$.

\begin{figure}[htbp]
        \centering
        \includegraphics[width=\onecol]{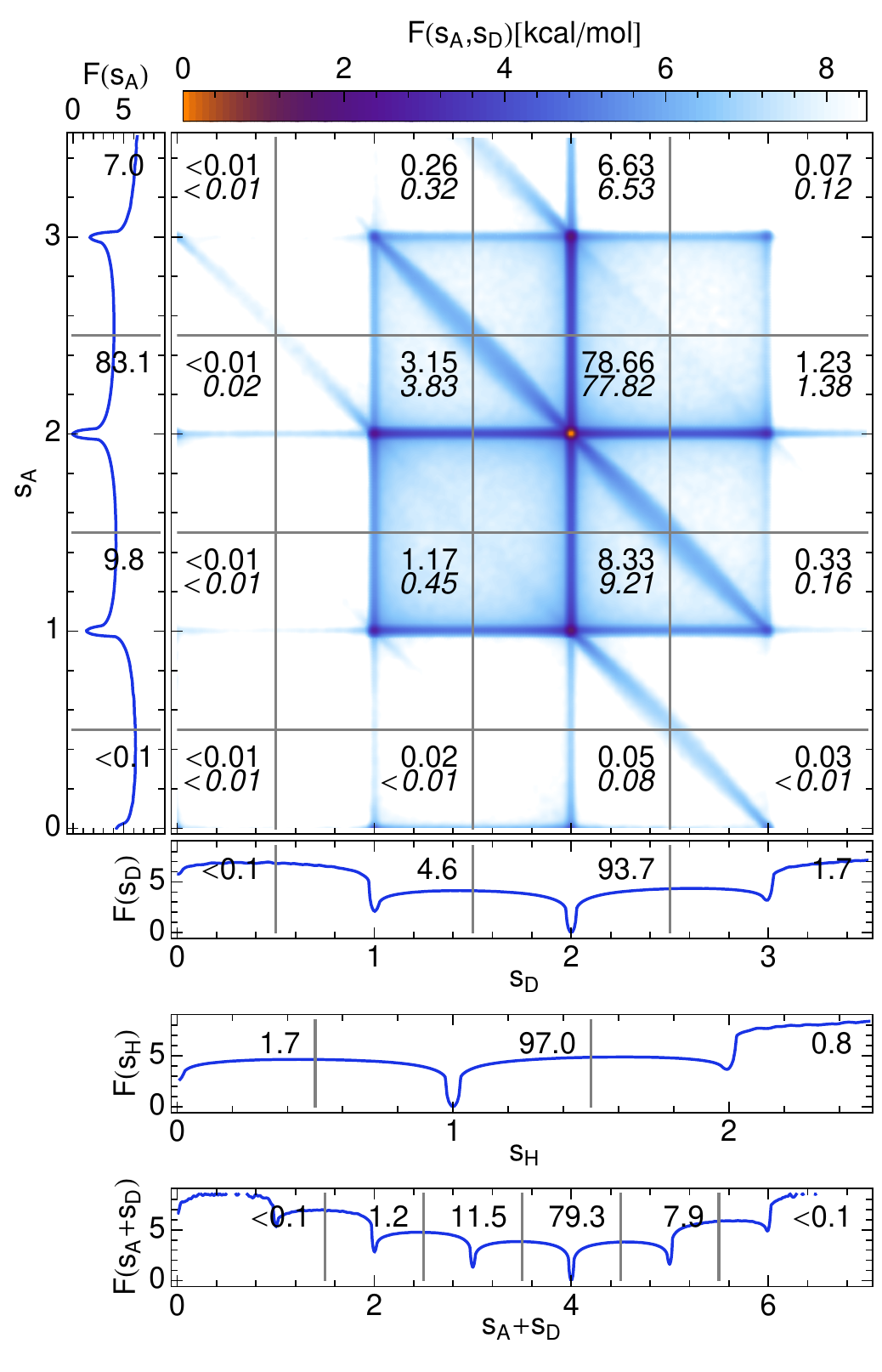}
        \caption{\label{fig:h2o-blyp-nqe}  Hydrogen-bond counts statistics for a PIGLET 
simulation of BLYP water at room temperature. See the caption of Figure~\ref{fig:h2o-tip4p-cls}
for a detailed explanation of the plots.}
\end{figure}

\paragraph{Quantum \emph{ab initio} water.} Finally, we considered a simulation that used the PIGLET technique to introduce nuclear 
quantum effects\cite{ceri-mano12prl} on top of a first-principles description of the 
electronic structure, analogous to the one used for the classical trajectory described above.
To achieve convergence of quantum properties, 6 beads were used together with a 
custom-tailored generalized Langevin equation thermostat,\cite{GLE4MD} 
as implemented in the i-PI Python interface.\cite{ceri+14cpc}
The overall statistics of the HB network (Figure~\ref{fig:h2o-blyp-nqe}) 
are not dramatically changed by nuclear quantum effects,
that only enhance marginally the probability of distorted configurations, with a decrease of ideal 
$(s_\text{D}\approx 2, s_\text{A}\approx 2)$ configurations and an increase of bifurcated hydrogen 
bonds. These are not however substantial changes, and are in part due to the fact that 
quantum fluctuations make the PAMM definition of $s_\text{DHA}$ less clear-cut than in the 
classical case.

Defining HBs with a method such as PAMM, that does not make any assumption on 
the covalent bonds present in the system, is particularly convenient in 
a context such as the present one. Extreme quantum fluctuations
of protons along the hydrogen bond lead to transient formal autolysis events, 
where the hydrogen atom detaches from the donor atom and reaches out to be 
closer to the acceptor oxygen.\cite{ceri+13pnas}
From a structural standpoint, PAMM does not recognize a distinct cluster corresponding to
these distorted configurations, but rather a continuum of structures. Starting from a \HB{O$_1$}{O$_2$} 
bond where the hydrogen atom is covalently bound to O$_1$, one goes smoothly through
distorted donated hydrogen bond to a configuration that is formally classified as 
a distorted bond \emph{accepted} by O$_1$ and \emph{donated} by O$_2$. 
These fluctuations conserve the total number of HBs between the two oxygen atoms, and only
change their character from donated to accepted. These excursions are apparent in the joint 
probability distribution of $s_\text{D}$ and $s_\text{A}$, where they show up as regions 
with higher probability extending diagonally between two near-integer $(s_\text{D},s_\text{A})$
regions.

\subsection*{Liquid ammonia}

As a final example, we considered liquid ammonia at 180K and ambient pressure. 
Configurations were kindly provided by Joshua More and David Manolopoulos.\cite{nh3unpub} 
Simulations were performed including nuclear quantum effects by path integral
molecular dynamics, using 12 beads and PIGLET\cite{ceri-mano12prl,GLE4MD} as
implemented in i-PI.\cite{ceri+14cpc} Quantum Espresso\cite{PWSCF} was used as the 
force back-end, with a PBE exchange-correlation functional\cite{perd+96prl}
and ultra-soft pseudo-potentials.\cite{vand90prb} The simulation box contained
32 molecules, and trajectories were performed for 10ps at constant, experimental
density, with the first 2~ps discarded for equilibration.

\begin{figure}[htbp]
        \centering
        \includegraphics[width=\onecol]{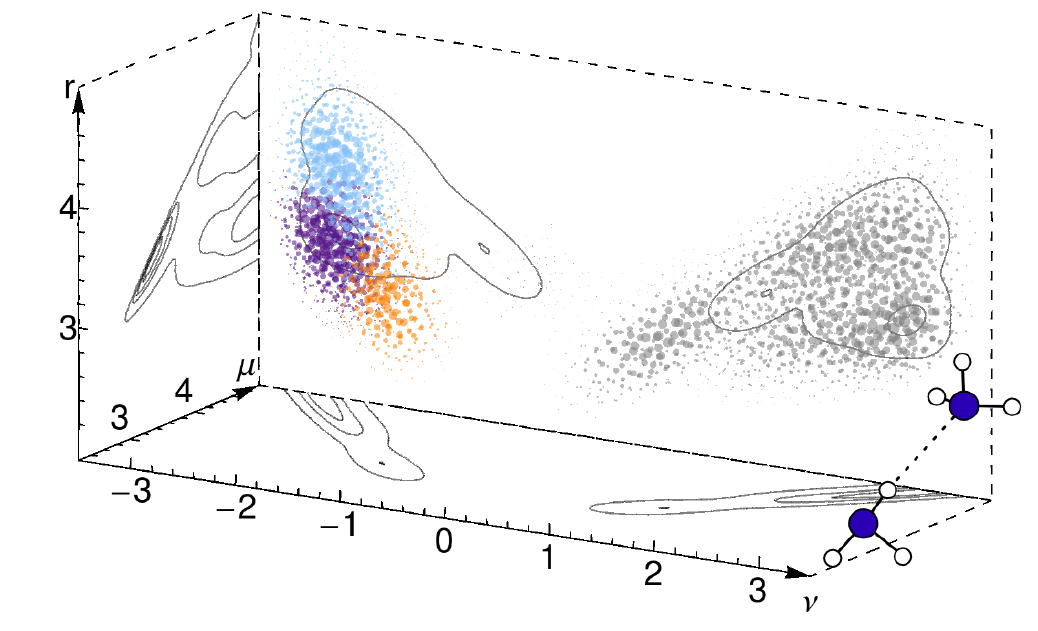}
        \caption{\label{fig:ammonia-3d}Distribution of $(\nu,\mu,r)$ configurations for \HB{O}{O$'$}
        in a simulation of liquid BLYP ammonia at 180K. Size and opacity of points correspond to the KDE of 
        $P(\mathbf{y})$, and colors indicate the cluster each grid point has been assigned to.
        Clusters with $\nu>0$ have not been colored, but have been correctly identified by PAMM. }
\end{figure}

Ammonia is a less-structured liquid than water, with weaker hydrogen bonds, as it is already
apparent from the probability density shown in Figure~\ref{fig:ammonia-3d}. Clusters are barely 
recognizable when using  a two-dimensional $(\nu,r)$ representation, which was instead 
capable of characterizing the HB in all the other cases we considered. 
Clustering is much more evident using the three-dimensional
$(\nu,\mu,r)$ description, and PAMM identifies clearly a range of values that can be 
ascribed to hydrogen-bonded configurations.
We expect the observation that higher-dimensional descriptors offer increased discriminating
power to be general, and provide a strategy to resolve weakly structured systems.
However, as the dimensionality is increased it becomes more difficult
to converge the probability distribution, so longer simulations are needed and
PAMM becomes more sensitive to the parameters of the procedure.
As it is the case for oxygen in H$_2$O, nitrogen atoms act both as 
donors and acceptors, leading to a symmetric structure for $P(\nu,\mu,r)$.

\begin{figure}[htbp]
        \centering
        \includegraphics[width=\onecol]{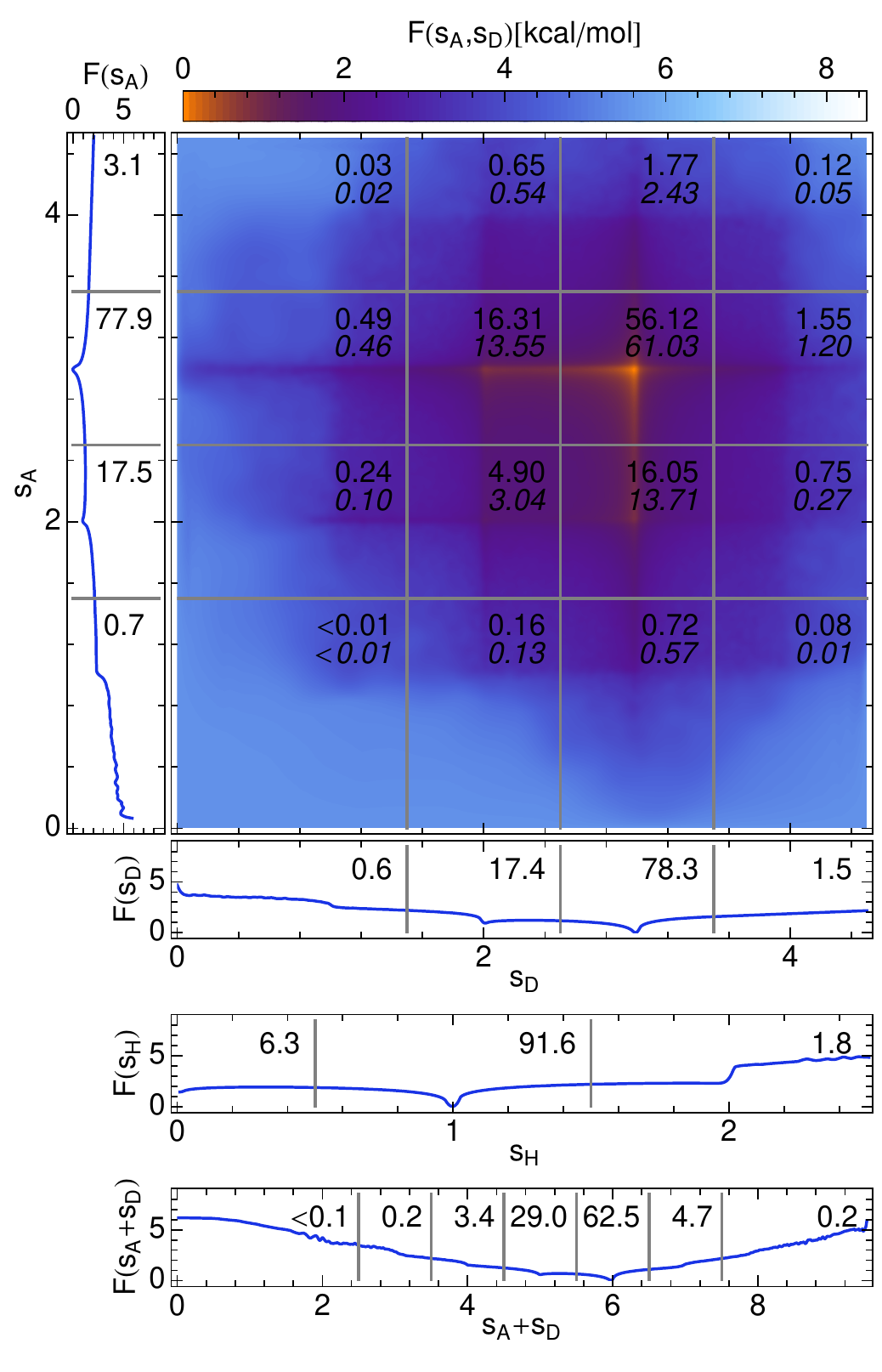}
        \caption{\label{fig:ammonia}  Hydrogen-bond counts statistics for a PIGLET 
simulation of BLYP liquid ammonia at 180K. See the caption of Figure~\ref{fig:h2o-tip4p-cls}
for a detailed explanation of the plots.}

\end{figure}

Figure~\ref{fig:ammonia} shows the analysis of the distribution of $s_\text{D}$, $s_\text{A}$ 
and $s_\text{H}$. Despite the low temperature,  the weaker and less directional HBs 
result in a less clear-cut  distribution of HB environments. 
A little more than 50\%{} of the nitrogen atoms receive and donate 3 HBs -- 
the ideal HB pattern that is observed in the solid phases of ammonia. Even though the 
simulation included nuclear quantum effects, there is no trace of the diagonal patterns
that are a manifestation of extreme proton excursions along the hydrogen bond.
Molecules maintain strictly their chemical integrity, and from a structural standpoint
these weak HBs appear to have a purely electrostatic character. 

\section*{Conclusions}

The probabilistic analysis of molecular motifs algorithm we have introduced attempts to 
reproduce in an automatic, data-driven manner the process of recognizing recurring 
structural patterns that is behind our intuitive understanding of chemical bonding. 
It first performs a non-parametric partitioning of the probability distribution that
characterizes the arrangement of a selected group of atoms in an atomistic simulation
of a material or a compound. It then proceeds to model the clusters with a Gaussian
mixture model that provides a natural, fuzzy definition of a chemical entity in terms 
of a smoothly varying posterior probability. 

We demonstrate the effectiveness of this approach by applying PAMM to recognize an ubiquitous
but hard to define entity as the hydrogen bond in a variety of different contexts. 
For each  \HB{D}{A} triplet of atoms, PAMM automatically identifies an appropriate range 
of structural parameters that provides an unbiased, agnostic definition of what constitutes
a hydrogen bond for a given set of atoms and a particular atomistic model. In the case of an empirical
forcefield model of solvated alanine dipeptide it identifies three (very different) ranges
of configurations that qualify as a distinct, recurring patterns for the carbonyl oxygen
accepting a HB from water, for the ammide nitrogen donating a HB to water, and for a 
hypothetical weak HB involving C$_\alpha$ atoms and water oxygens. In the latter case, the 
presence of a distinct feature in the probability distribution is probably an indirect 
effect of the structural correlations in the water HB network, of the HBs between water
and the electro-negative atoms in alanine dipeptide and of the rigidity of the molecular 
backbone. 

We then assessed the behavior of PAMM when performing a more detailed 
analysis of hydrogen bonding in water, comparing an empirical water model and 
a first-principles, density functional model of water with and without a description
of the quantum nature of nuclei. We introduced a compact representation of the 
hydrogen-bonding properties of water molecules in terms of the total number
of accepted and donated HBs, that arises naturally because PAMM identifies
hydrogen-bonded configurations in terms of a smoothly varying HB count function. 
We demonstrated that these hydrogen bond counts can be used to study the 
dynamics of the hydrogen bond network, giving results that are fully compatible
with well-established definitions of the hydrogen bond, \blue{and to identify
coordination defects in the otherwise ideal HB network of ice Ih}.
This analysis also highlights the presence of characteristic features that 
are a signature of extreme excursions of protons along the hydrogen bond observed
with a quantum description of nuclei.
Finally, we discussed liquid ammonia as an example of a weakly hydrogen-bonded 
system, that shows a much less clear-cut partitioning of the probability distribution 
and a more varied ensemble of hydrogen-bonding molecular environments. 

In all of these cases our algorithm provides an adaptive approach to define
the hydrogen bond in a unique and unbiased way, that uses only structural information
and that can be easily exploited to recognize correlation between the HB patterns
involving a given molecule. Even though we have used hydrogen bonding as a
representative benchmark, application of the PAMM algorithm is by no means limited to 
this example. It could be used to recognize complex structural patterns in a variety
of materials and compounds, and be easily applied to bias molecular 
simulations to accelerate the interconversion between different 
(meta)stable atomic configurations.

\section*{Acknowledgments}

The authors would like to thank Gareth Tribello, Federico Giberti and Fabio Pietrucci
for a careful reading of the manuscript and useful discussion. Joshua More and David Manolopoulos 
are thanked for sharing the ammonia configurations with us. Funding from the
National Center for Computational Design and Discovery of Novel Materials MARVEL
is gratefully acknowledged, as well as computer time from the Swiss National Supercomputing Center (project s466).

\appendix

\section{PAMM library and utilities}\label{app:pamm}

In the Supplementary Material we provide the source code of a simple Fortran90 
library to perform PAMM analysis, and a tool to apply it to the case of 
the hydrogen bond~\cite{si}. Updated versions of the package will be made available on
\url{http://epfl-cosmo.github.io}.
 The workflow for using this software follows closely the procedure
introduced in Section~\ref{sec:methods}. Starting from a preliminary simulation,
one should first extract representative descriptors for the atom groups that are 
being studied. Here, one can use the \emph{hbpamm} utility to process a set of
atomic configurations in \emph{xyz} format, to generate a list of $(\nu,\mu,r)$ 
triplets. Then, the \emph{pamm} tool can use this data to generate a probabilistic 
model that describes the range of structural parameters associated with the 
hydrogen bond. This piece of code is completely general, and can be used to 
process any sort of structural data. Finally, one should use once more 
\emph{hbpamm} (or an analogous utility adapted to a different kind of
structure) to post-process the trajectory, extracting $s_\text{A}$, $s_\text{D}$, 
$s_\text{H}$ for the different atoms considered.


\begin{thebibliography}{67}%
\makeatletter
\providecommand \@ifxundefined [1]{%
 \@ifx{#1\undefined}
}%
\providecommand \@ifnum [1]{%
 \ifnum #1\expandafter \@firstoftwo
 \else \expandafter \@secondoftwo
 \fi
}%
\providecommand \@ifx [1]{%
 \ifx #1\expandafter \@firstoftwo
 \else \expandafter \@secondoftwo
 \fi
}%
\providecommand \natexlab [1]{#1}%
\providecommand \enquote  [1]{``#1''}%
\providecommand \bibnamefont  [1]{#1}%
\providecommand \bibfnamefont [1]{#1}%
\providecommand \citenamefont [1]{#1}%
\providecommand \href@noop [0]{\@secondoftwo}%
\providecommand \href [0]{\begingroup \@sanitize@url \@href}%
\providecommand \@href[1]{\@@startlink{#1}\@@href}%
\providecommand \@@href[1]{\endgroup#1\@@endlink}%
\providecommand \@sanitize@url [0]{\catcode `\\12\catcode `\$12\catcode
  `\&12\catcode `\#12\catcode `\^12\catcode `\_12\catcode `\%12\relax}%
\providecommand \@@startlink[1]{}%
\providecommand \@@endlink[0]{}%
\providecommand \url  [0]{\begingroup\@sanitize@url \@url }%
\providecommand \@url [1]{\endgroup\@href {#1}{\urlprefix }}%
\providecommand \urlprefix  [0]{URL }%
\providecommand \Eprint [0]{\href }%
\providecommand \doibase [0]{http://dx.doi.org/}%
\providecommand \selectlanguage [0]{\@gobble}%
\providecommand \bibinfo  [0]{\@secondoftwo}%
\providecommand \bibfield  [0]{\@secondoftwo}%
\providecommand \translation [1]{[#1]}%
\providecommand \BibitemOpen [0]{}%
\providecommand \bibitemStop [0]{}%
\providecommand \bibitemNoStop [0]{.\EOS\space}%
\providecommand \EOS [0]{\spacefactor3000\relax}%
\providecommand \BibitemShut  [1]{\csname bibitem#1\endcsname}%
\let\auto@bib@innerbib\@empty
\bibitem [{\citenamefont {Pauling}(1960)}]{paul1960book}%
  \BibitemOpen
  \bibfield  {author} {\bibinfo {author} {\bibfnamefont {L.}~\bibnamefont
  {Pauling}},\ }\href@noop {} {\emph {\bibinfo {title} {The nature of the
  chemical bond and the structure of molecules and crystals: an introduction to
  modern structural chemistry}}},\ Vol.~\bibinfo {volume} {18}\ (\bibinfo
  {publisher} {Cornell University Press},\ \bibinfo {year} {1960})\BibitemShut
  {NoStop}%
\bibitem [{\citenamefont {Car}\ and\ \citenamefont
  {Parrinello}(1985)}]{car-parr85prl}%
  \BibitemOpen
  \bibfield  {author} {\bibinfo {author} {\bibfnamefont {R.}~\bibnamefont
  {Car}}\ and\ \bibinfo {author} {\bibfnamefont {M.}~\bibnamefont
  {Parrinello}},\ }\href@noop {} {\bibfield  {journal} {\bibinfo  {journal}
  {Phys. Rev. Lett.}\ }\textbf {\bibinfo {volume} {55}},\ \bibinfo {pages}
  {2471} (\bibinfo {year} {1985})}\BibitemShut {NoStop}%
\bibitem [{\citenamefont {Schaefer~III}(2012)}]{scha-henr12book}%
  \BibitemOpen
  \bibfield  {author} {\bibinfo {author} {\bibfnamefont {H.~F.}\ \bibnamefont
  {Schaefer~III}},\ }\href@noop {} {\emph {\bibinfo {title} {Quantum chemistry:
  the development of ab initio methods in molecular electronic structure
  theory}}}\ (\bibinfo  {publisher} {Courier Dover Publications},\ \bibinfo
  {year} {2012})\BibitemShut {NoStop}%
\bibitem [{\citenamefont {MacKerell}\ \emph {et~al.}(1998)\citenamefont
  {MacKerell}, \citenamefont {Bashford}, \citenamefont {Bellott}, \citenamefont
  {Dunbrack}, \citenamefont {Evanseck}, \citenamefont {Field}, \citenamefont
  {Fischer}, \citenamefont {Gao}, \citenamefont {Guo}, \citenamefont {Ha},\
  and\ \citenamefont {{others}}}]{mack+98JPhysChemB}%
  \BibitemOpen
  \bibfield  {author} {\bibinfo {author} {\bibfnamefont {A.~D.}\ \bibnamefont
  {MacKerell}}, \bibinfo {author} {\bibfnamefont {D.}~\bibnamefont {Bashford}},
  \bibinfo {author} {\bibfnamefont {M.}~\bibnamefont {Bellott}}, \bibinfo
  {author} {\bibfnamefont {R.~L.}\ \bibnamefont {Dunbrack}}, \bibinfo {author}
  {\bibfnamefont {J.~D.}\ \bibnamefont {Evanseck}}, \bibinfo {author}
  {\bibfnamefont {M.~J.}\ \bibnamefont {Field}}, \bibinfo {author}
  {\bibfnamefont {S.}~\bibnamefont {Fischer}}, \bibinfo {author} {\bibfnamefont
  {J.}~\bibnamefont {Gao}}, \bibinfo {author} {\bibfnamefont {H.}~\bibnamefont
  {Guo}}, \bibinfo {author} {\bibfnamefont {S.~a.}\ \bibnamefont {Ha}}, \ and\
  \bibinfo {author} {\bibnamefont {{others}}},\ }\href@noop {} {\bibfield
  {journal} {\bibinfo  {journal} {The Journal of Physical Chemistry B}\
  }\textbf {\bibinfo {volume} {102}},\ \bibinfo {pages} {3586} (\bibinfo {year}
  {1998})}\BibitemShut {NoStop}%
\bibitem [{\citenamefont {Behler}\ and\ \citenamefont
  {Parrinello}(2007)}]{behl-parr07prl}%
  \BibitemOpen
  \bibfield  {author} {\bibinfo {author} {\bibfnamefont {J.}~\bibnamefont
  {Behler}}\ and\ \bibinfo {author} {\bibfnamefont {M.}~\bibnamefont
  {Parrinello}},\ }\href@noop {} {\bibfield  {journal} {\bibinfo  {journal}
  {Phys. Rev. Lett.}\ }\textbf {\bibinfo {volume} {98}},\ \bibinfo {pages}
  {146401} (\bibinfo {year} {2007})}\BibitemShut {NoStop}%
\bibitem [{\citenamefont {Bart\'{o}k}\ \emph {et~al.}(2010)\citenamefont
  {Bart\'{o}k}, \citenamefont {Payne}, \citenamefont {Kondor},\ and\
  \citenamefont {Cs\'{a}nyi}}]{bart+10prl}%
  \BibitemOpen
  \bibfield  {author} {\bibinfo {author} {\bibfnamefont {A.~P.}\ \bibnamefont
  {Bart\'{o}k}}, \bibinfo {author} {\bibfnamefont {M.~C.}\ \bibnamefont
  {Payne}}, \bibinfo {author} {\bibfnamefont {R.}~\bibnamefont {Kondor}}, \
  and\ \bibinfo {author} {\bibfnamefont {G.}~\bibnamefont {Cs\'{a}nyi}},\
  }\href@noop {} {\bibfield  {journal} {\bibinfo  {journal} {Phys. Rev. Lett.}\
  }\textbf {\bibinfo {volume} {104}},\ \bibinfo {pages} {136403} (\bibinfo
  {year} {2010})}\BibitemShut {NoStop}%
\bibitem [{\citenamefont {Pauling}(1928)}]{paul1928PNAS}%
  \BibitemOpen
  \bibfield  {author} {\bibinfo {author} {\bibfnamefont {L.}~\bibnamefont
  {Pauling}},\ }\href@noop {} {\bibfield  {journal} {\bibinfo  {journal}
  {Proceedings of the National Academy of Sciences of the United States of
  America}\ }\textbf {\bibinfo {volume} {14}},\ \bibinfo {pages} {359}
  (\bibinfo {year} {1928})}\BibitemShut {NoStop}%
\bibitem [{\citenamefont {Pauling}\ \emph {et~al.}(1951)\citenamefont
  {Pauling}, \citenamefont {Corey},\ and\ \citenamefont
  {Branson}}]{paul51PNAS}%
  \BibitemOpen
  \bibfield  {author} {\bibinfo {author} {\bibfnamefont {L.}~\bibnamefont
  {Pauling}}, \bibinfo {author} {\bibfnamefont {R.~B.}\ \bibnamefont {Corey}},
  \ and\ \bibinfo {author} {\bibfnamefont {H.~R.}\ \bibnamefont {Branson}},\
  }\href@noop {} {\bibfield  {journal} {\bibinfo  {journal} {Proceedings of the
  National Academy of Sciences}\ }\textbf {\bibinfo {volume} {37}},\ \bibinfo
  {pages} {205} (\bibinfo {year} {1951})}\BibitemShut {NoStop}%
\bibitem [{\citenamefont {Guo}\ and\ \citenamefont
  {Karplus}(1994)}]{hong-karp94JPhysChem}%
  \BibitemOpen
  \bibfield  {author} {\bibinfo {author} {\bibfnamefont {H.}~\bibnamefont
  {Guo}}\ and\ \bibinfo {author} {\bibfnamefont {M.}~\bibnamefont {Karplus}},\
  }\href@noop {} {\bibfield  {journal} {\bibinfo  {journal} {The Journal of
  Physical Chemistry}\ }\textbf {\bibinfo {volume} {98}},\ \bibinfo {pages}
  {7104} (\bibinfo {year} {1994})}\BibitemShut {NoStop}%
\bibitem [{\citenamefont {Jeffrey}(1997)}]{jeff97book}%
  \BibitemOpen
  \bibfield  {author} {\bibinfo {author} {\bibfnamefont {G.~A.}\ \bibnamefont
  {Jeffrey}},\ }\href@noop {} {\emph {\bibinfo {title} {An introduction to
  hydrogen bonding}}},\ Vol.~\bibinfo {volume} {12}\ (\bibinfo  {publisher}
  {Oxford university press New York},\ \bibinfo {year} {1997})\BibitemShut
  {NoStop}%
\bibitem [{\citenamefont {Desiraju}\ and\ \citenamefont
  {Steiner}(2001)}]{desi-stei01book}%
  \BibitemOpen
  \bibfield  {author} {\bibinfo {author} {\bibfnamefont {G.~R.}\ \bibnamefont
  {Desiraju}}\ and\ \bibinfo {author} {\bibfnamefont {T.}~\bibnamefont
  {Steiner}},\ }\href@noop {} {\emph {\bibinfo {title} {Weak hydrogen Bond}}}\
  (\bibinfo  {publisher} {Oxford University Press New York},\ \bibinfo {year}
  {2001})\BibitemShut {NoStop}%
\bibitem [{\citenamefont {Lin}\ \emph {et~al.}(2011)\citenamefont {Lin},
  \citenamefont {Morrone},\ and\ \citenamefont {Car}}]{lin+11jsp}%
  \BibitemOpen
  \bibfield  {author} {\bibinfo {author} {\bibfnamefont {L.}~\bibnamefont
  {Lin}}, \bibinfo {author} {\bibfnamefont {J.~a.}\ \bibnamefont {Morrone}}, \
  and\ \bibinfo {author} {\bibfnamefont {R.}~\bibnamefont {Car}},\ }\href@noop
  {} {\bibfield  {journal} {\bibinfo  {journal} {J. Stat. Phys.}\ }\textbf
  {\bibinfo {volume} {145}},\ \bibinfo {pages} {365} (\bibinfo {year}
  {2011})}\BibitemShut {NoStop}%
\bibitem [{\citenamefont {McKenzie}\ \emph {et~al.}(2014)\citenamefont
  {McKenzie}, \citenamefont {Bekker}, \citenamefont {Athokpam},\ and\
  \citenamefont {Ramesh}}]{mcke+14jcp}%
  \BibitemOpen
  \bibfield  {author} {\bibinfo {author} {\bibfnamefont {R.~H.}\ \bibnamefont
  {McKenzie}}, \bibinfo {author} {\bibfnamefont {C.}~\bibnamefont {Bekker}},
  \bibinfo {author} {\bibfnamefont {B.}~\bibnamefont {Athokpam}}, \ and\
  \bibinfo {author} {\bibfnamefont {S.~G.}\ \bibnamefont {Ramesh}},\
  }\href@noop {} {\bibfield  {journal} {\bibinfo  {journal} {J. Chem. Phys.}\
  }\textbf {\bibinfo {volume} {140}},\ \bibinfo {pages} {174508} (\bibinfo
  {year} {2014})}\BibitemShut {NoStop}%
\bibitem [{\citenamefont {Weinhold}\ and\ \citenamefont
  {Klein}(2012)}]{wein-klei2012MolPhys}%
  \BibitemOpen
  \bibfield  {author} {\bibinfo {author} {\bibfnamefont {F.}~\bibnamefont
  {Weinhold}}\ and\ \bibinfo {author} {\bibfnamefont {R.~A.}\ \bibnamefont
  {Klein}},\ }\href@noop {} {\bibfield  {journal} {\bibinfo  {journal}
  {Molecular Physics}\ }\textbf {\bibinfo {volume} {110}},\ \bibinfo {pages}
  {565} (\bibinfo {year} {2012})}\BibitemShut {NoStop}%
\bibitem [{\citenamefont {Wendler}\ \emph {et~al.}(2010)\citenamefont
  {Wendler}, \citenamefont {Thar}, \citenamefont {Zahn},\ and\ \citenamefont
  {Kirchner}}]{wend+10jpca}%
  \BibitemOpen
  \bibfield  {author} {\bibinfo {author} {\bibfnamefont {K.}~\bibnamefont
  {Wendler}}, \bibinfo {author} {\bibfnamefont {J.}~\bibnamefont {Thar}},
  \bibinfo {author} {\bibfnamefont {S.}~\bibnamefont {Zahn}}, \ and\ \bibinfo
  {author} {\bibfnamefont {B.}~\bibnamefont {Kirchner}},\ }\href@noop {}
  {\bibfield  {journal} {\bibinfo  {journal} {J. Phys. Chem. A}\ }\textbf
  {\bibinfo {volume} {114}},\ \bibinfo {pages} {9529} (\bibinfo {year}
  {2010})}\BibitemShut {NoStop}%
\bibitem [{\citenamefont {Azar}\ and\ \citenamefont
  {Head-Gordon}(2012)}]{azarhead12JChemPhys}%
  \BibitemOpen
  \bibfield  {author} {\bibinfo {author} {\bibfnamefont {R.~J.}\ \bibnamefont
  {Azar}}\ and\ \bibinfo {author} {\bibfnamefont {M.}~\bibnamefont
  {Head-Gordon}},\ }\href {\doibase 10.1063/1.3674992} {\bibfield  {journal}
  {\bibinfo  {journal} {The Journal of Chemical Physics}\ }\textbf {\bibinfo
  {volume} {136}},\ \bibinfo {pages} {024103} (\bibinfo {year}
  {2012})}\BibitemShut {NoStop}%
\bibitem [{\citenamefont {Weinhold}\ and\ \citenamefont
  {Klein}(2014)}]{wein-klei14ChemEducResPract}%
  \BibitemOpen
  \bibfield  {author} {\bibinfo {author} {\bibfnamefont {F.}~\bibnamefont
  {Weinhold}}\ and\ \bibinfo {author} {\bibfnamefont {R.~A.}\ \bibnamefont
  {Klein}},\ }\href {\doibase 10.1039/c4rp00030g} {\bibfield  {journal}
  {\bibinfo  {journal} {Chemistry Education Research and Practice}\ } (\bibinfo
  {year} {2014}),\ 10.1039/c4rp00030g}\BibitemShut {NoStop}%
\bibitem [{\citenamefont {K\"{u}hne}\ and\ \citenamefont
  {Khaliullin}(2014)}]{kuhn-khal14jacs}%
  \BibitemOpen
  \bibfield  {author} {\bibinfo {author} {\bibfnamefont {T.~D.}\ \bibnamefont
  {K\"{u}hne}}\ and\ \bibinfo {author} {\bibfnamefont {R.~Z.}\ \bibnamefont
  {Khaliullin}},\ }\href@noop {} {\bibfield  {journal} {\bibinfo  {journal} {J.
  Am. Chem. Soc.}\ }\textbf {\bibinfo {volume} {136}},\ \bibinfo {pages} {3395}
  (\bibinfo {year} {2014})}\BibitemShut {NoStop}%
\bibitem [{\citenamefont {Vedaldi}\ and\ \citenamefont
  {Soatto}(2008)}]{veda-soat08chapterbook}%
  \BibitemOpen
  \bibfield  {author} {\bibinfo {author} {\bibfnamefont {A.}~\bibnamefont
  {Vedaldi}}\ and\ \bibinfo {author} {\bibfnamefont {S.}~\bibnamefont
  {Soatto}},\ }in\ \href
  {http://link.springer.com/chapter/10.1007/978-3-540-88693-8_52} {\emph
  {\bibinfo {booktitle} {Computer {Vision-ECCV} 2008}}}\ (\bibinfo  {publisher}
  {Springer},\ \bibinfo {year} {2008})\ pp.\ \bibinfo {pages}
  {705--718}\BibitemShut {NoStop}%
\bibitem [{si()}]{si}%
  \BibitemOpen
  \href@noop {} {}\bibinfo {howpublished} {See Supplementary Material Document
  No. xxxxxx for a detailed discussion of the stability of the algorithm, a
  library of tools to apply PAMM, and the input files to reproduce the
  simulations described in the text.}\BibitemShut {Stop}%
\bibitem [{\citenamefont {Gallet}\ and\ \citenamefont
  {Pietrucci}(2013)}]{gall-piet13jcp}%
  \BibitemOpen
  \bibfield  {author} {\bibinfo {author} {\bibfnamefont {G.}~\bibnamefont
  {Gallet}}\ and\ \bibinfo {author} {\bibfnamefont {F.}~\bibnamefont
  {Pietrucci}},\ }\href@noop {} {\bibfield  {journal} {\bibinfo  {journal} {J.
  Chem. Phys.}\ }\textbf {\bibinfo {volume} {139}},\ \bibinfo {pages} {074101}
  (\bibinfo {year} {2013})}\BibitemShut {NoStop}%
\bibitem [{Note1()}]{Note1}%
  \BibitemOpen
  \bibinfo {note} {In practice, one can almost invariably introduce a cutoff to
  avoid considering tuples that involve atoms that are very far away from each
  other, and hence clearly unrelated.}\BibitemShut {Stop}%
\bibitem [{\citenamefont {Silverman}(1986)}]{silv86book}%
  \BibitemOpen
  \bibfield  {author} {\bibinfo {author} {\bibfnamefont {B.~W.}\ \bibnamefont
  {Silverman}},\ }\href@noop {} {\emph {\bibinfo {title} {Density estimation
  for statistics and data analysis}}},\ Vol.~\bibinfo {volume} {26}\ (\bibinfo
  {publisher} {CRC press},\ \bibinfo {year} {1986})\BibitemShut {NoStop}%
\bibitem [{\citenamefont {Eldar}\ \emph {et~al.}(1997)\citenamefont {Eldar},
  \citenamefont {Lindenbaum}, \citenamefont {Porat},\ and\ \citenamefont
  {Zeevi}}]{elda+97ieee}%
  \BibitemOpen
  \bibfield  {author} {\bibinfo {author} {\bibfnamefont {Y.}~\bibnamefont
  {Eldar}}, \bibinfo {author} {\bibfnamefont {M.}~\bibnamefont {Lindenbaum}},
  \bibinfo {author} {\bibfnamefont {M.}~\bibnamefont {Porat}}, \ and\ \bibinfo
  {author} {\bibfnamefont {Y.~Y.}\ \bibnamefont {Zeevi}},\ }\href@noop {}
  {\bibfield  {journal} {\bibinfo  {journal} {IEEE Trans. Image Process.}\
  }\textbf {\bibinfo {volume} {6}},\ \bibinfo {pages} {1305} (\bibinfo {year}
  {1997})}\BibitemShut {NoStop}%
\bibitem [{Note2()}]{Note2}%
  \BibitemOpen
  \bibinfo {note} {As a trivial example of such weights, consider the $1/r^2$
  normalization that is used to define the radial distribution function between
  a pair of atoms based on the histogram of the pairwise distances
  $r$.}\BibitemShut {Stop}%
\bibitem [{\citenamefont {Comaniciu}\ and\ \citenamefont
  {Meer}(2002)}]{coma-meer02ieee}%
  \BibitemOpen
  \bibfield  {author} {\bibinfo {author} {\bibfnamefont {D.}~\bibnamefont
  {Comaniciu}}\ and\ \bibinfo {author} {\bibfnamefont {P.}~\bibnamefont
  {Meer}},\ }\href@noop {} {\bibfield  {journal} {\bibinfo  {journal} {IEEE
  Trans. Pattern Anal. Mach. Intell.}\ }\textbf {\bibinfo {volume} {24}},\
  \bibinfo {pages} {603} (\bibinfo {year} {2002})}\BibitemShut {NoStop}%
\bibitem [{Note3()}]{Note3}%
  \BibitemOpen
  \bibinfo {note} {Quick shift is not particularly sensitive to the choice of
  $\lambda $ We found it convenient to set $\lambda =5\left <\sigma _i\right
  >$, where the $\sigma _i$ are the kernel widths associated with the grid
  points. The SM~\cite {si}.}\BibitemShut {Stop}%
\bibitem [{Note4()}]{Note4}%
  \BibitemOpen
  \bibinfo {note} {Quick shift is a \protect \emph {medoid} method, and as such
  the modes it finds are bound to be grid points. This means that the
  resolution of the grid impacts the definition of the cluster centers. In
  practice, this is easily solved by quickly optimizing the positions of the
  points $\protect \mathbf {z}_k$ using mean shift.}\BibitemShut {Stop}%
\bibitem [{\citenamefont {Press}(2007)}]{press07book}%
  \BibitemOpen
  \bibfield  {author} {\bibinfo {author} {\bibfnamefont {W.~H.}\ \bibnamefont
  {Press}},\ }\href@noop {} {\emph {\bibinfo {title} {{Numerical Recipes: The
  art of scientific computing}}}}\ (\bibinfo  {publisher} {Cambridge University
  Press},\ \bibinfo {year} {2007})\BibitemShut {NoStop}%
\bibitem [{\citenamefont {Reynolds}(2009)}]{reyn09book}%
  \BibitemOpen
  \bibfield  {author} {\bibinfo {author} {\bibfnamefont {D.}~\bibnamefont
  {Reynolds}},\ }\href@noop {} {\bibfield  {journal} {\bibinfo  {journal}
  {Encyclopedia of Biometrics}\ ,\ \bibinfo {pages} {659}} (\bibinfo {year}
  {2009})}\BibitemShut {NoStop}%
\bibitem [{\citenamefont {Maragakis}\ \emph {et~al.}(2009)\citenamefont
  {Maragakis}, \citenamefont {van~der Vaart},\ and\ \citenamefont
  {Karplus}}]{mara+09jpcb}%
  \BibitemOpen
  \bibfield  {author} {\bibinfo {author} {\bibfnamefont {P.}~\bibnamefont
  {Maragakis}}, \bibinfo {author} {\bibfnamefont {A.}~\bibnamefont {van~der
  Vaart}}, \ and\ \bibinfo {author} {\bibfnamefont {M.}~\bibnamefont
  {Karplus}},\ }\href@noop {} {\bibfield  {journal} {\bibinfo  {journal} {J.
  Phys. Chem. B}\ }\textbf {\bibinfo {volume} {113}},\ \bibinfo {pages} {4664}
  (\bibinfo {year} {2009})}\BibitemShut {NoStop}%
\bibitem [{\citenamefont {Tribello}\ \emph {et~al.}(2010)\citenamefont
  {Tribello}, \citenamefont {Ceriotti},\ and\ \citenamefont
  {Parrinello}}]{trib+10pnas}%
  \BibitemOpen
  \bibfield  {author} {\bibinfo {author} {\bibfnamefont {G.~A.}\ \bibnamefont
  {Tribello}}, \bibinfo {author} {\bibfnamefont {M.}~\bibnamefont {Ceriotti}},
  \ and\ \bibinfo {author} {\bibfnamefont {M.}~\bibnamefont {Parrinello}},\
  }\href@noop {} {\bibfield  {journal} {\bibinfo  {journal} {Proc. Natl. Acad.
  Sci. USA U. S. A.}\ }\textbf {\bibinfo {volume} {107}},\ \bibinfo {pages}
  {17509} (\bibinfo {year} {2010})}\BibitemShut {NoStop}%
\bibitem [{\citenamefont {Pisani}\ \emph {et~al.}(2014)\citenamefont {Pisani},
  \citenamefont {Piro}, \citenamefont {Decherchi}, \citenamefont {Bottegoni},
  \citenamefont {Sona}, \citenamefont {Murino}, \citenamefont {Rocchia},\ and\
  \citenamefont {Cavalli}}]{pisa+14jctc}%
  \BibitemOpen
  \bibfield  {author} {\bibinfo {author} {\bibfnamefont {P.}~\bibnamefont
  {Pisani}}, \bibinfo {author} {\bibfnamefont {P.}~\bibnamefont {Piro}},
  \bibinfo {author} {\bibfnamefont {S.}~\bibnamefont {Decherchi}}, \bibinfo
  {author} {\bibfnamefont {G.}~\bibnamefont {Bottegoni}}, \bibinfo {author}
  {\bibfnamefont {D.}~\bibnamefont {Sona}}, \bibinfo {author} {\bibfnamefont
  {V.}~\bibnamefont {Murino}}, \bibinfo {author} {\bibfnamefont
  {W.}~\bibnamefont {Rocchia}}, \ and\ \bibinfo {author} {\bibfnamefont
  {A.}~\bibnamefont {Cavalli}},\ }\href@noop {} {\bibfield  {journal} {\bibinfo
   {journal} {J. Chem. Theory Comput.}\ }\textbf {\bibinfo {volume} {10}},\
  \bibinfo {pages} {2557} (\bibinfo {year} {2014})}\BibitemShut {NoStop}%
\bibitem [{\citenamefont {Laio}\ and\ \citenamefont
  {Parrinello}(2002)}]{laio-parr02pnas}%
  \BibitemOpen
  \bibfield  {author} {\bibinfo {author} {\bibfnamefont {A.}~\bibnamefont
  {Laio}}\ and\ \bibinfo {author} {\bibfnamefont {M.}~\bibnamefont
  {Parrinello}},\ }\href@noop {} {\bibfield  {journal} {\bibinfo  {journal}
  {Proc. Natl. Acad. Sci. USA}\ }\textbf {\bibinfo {volume} {99}},\ \bibinfo
  {pages} {12562} (\bibinfo {year} {2002})}\BibitemShut {NoStop}%
\bibitem [{\citenamefont {Pietrucci}\ and\ \citenamefont
  {Andreoni}(2011)}]{piet-andr11prl}%
  \BibitemOpen
  \bibfield  {author} {\bibinfo {author} {\bibfnamefont {F.}~\bibnamefont
  {Pietrucci}}\ and\ \bibinfo {author} {\bibfnamefont {W.}~\bibnamefont
  {Andreoni}},\ }\href@noop {} {\bibfield  {journal} {\bibinfo  {journal}
  {Phys. Rev. Lett.}\ }\textbf {\bibinfo {volume} {107}},\ \bibinfo {pages}
  {085504} (\bibinfo {year} {2011})}\BibitemShut {NoStop}%
\bibitem [{\citenamefont {Sadeghi}\ and\ \citenamefont
  {Ghasemi}(2013)}]{sade+13jcp}%
  \BibitemOpen
  \bibfield  {author} {\bibinfo {author} {\bibfnamefont {A.}~\bibnamefont
  {Sadeghi}}\ and\ \bibinfo {author} {\bibfnamefont {S.}~\bibnamefont
  {Ghasemi}},\ }\href@noop {} {\bibfield  {journal} {\bibinfo  {journal} {J.
  Chem. Phys.}\ }\textbf {\bibinfo {volume} {139}},\ \bibinfo {pages} {184118}
  (\bibinfo {year} {2013})}\BibitemShut {NoStop}%
\bibitem [{\citenamefont {Ferguson}\ \emph {et~al.}(2010)\citenamefont
  {Ferguson}, \citenamefont {Panagiotopoulos}, \citenamefont {Debenedetti},\
  and\ \citenamefont {Kevrekidis}}]{ferg+10pnas}%
  \BibitemOpen
  \bibfield  {author} {\bibinfo {author} {\bibfnamefont {A.~L.}\ \bibnamefont
  {Ferguson}}, \bibinfo {author} {\bibfnamefont {A.~Z.}\ \bibnamefont
  {Panagiotopoulos}}, \bibinfo {author} {\bibfnamefont {P.~G.}\ \bibnamefont
  {Debenedetti}}, \ and\ \bibinfo {author} {\bibfnamefont {I.~G.}\ \bibnamefont
  {Kevrekidis}},\ }\href@noop {} {\bibfield  {journal} {\bibinfo  {journal}
  {Proc. Natl. Acad. Sci. USA U. S. A.}\ }\textbf {\bibinfo {volume} {107}},\
  \bibinfo {pages} {13597} (\bibinfo {year} {2010})}\BibitemShut {NoStop}%
\bibitem [{\citenamefont {Tribello}\ \emph {et~al.}(2012)\citenamefont
  {Tribello}, \citenamefont {Ceriotti},\ and\ \citenamefont
  {Parrinello}}]{trib+12pnas}%
  \BibitemOpen
  \bibfield  {author} {\bibinfo {author} {\bibfnamefont {G.~A.}\ \bibnamefont
  {Tribello}}, \bibinfo {author} {\bibfnamefont {M.}~\bibnamefont {Ceriotti}},
  \ and\ \bibinfo {author} {\bibfnamefont {M.}~\bibnamefont {Parrinello}},\
  }\href@noop {} {\bibfield  {journal} {\bibinfo  {journal} {Proc. Natl. Acad.
  Sci. USA U. S. A.}\ }\textbf {\bibinfo {volume} {109}},\ \bibinfo {pages}
  {5196} (\bibinfo {year} {2012})}\BibitemShut {NoStop}%
\bibitem [{\citenamefont {Rohrdanz}\ \emph {et~al.}(2013)\citenamefont
  {Rohrdanz}, \citenamefont {Zheng},\ and\ \citenamefont
  {Clementi}}]{rohr+arpc03}%
  \BibitemOpen
  \bibfield  {author} {\bibinfo {author} {\bibfnamefont {M.~a.}\ \bibnamefont
  {Rohrdanz}}, \bibinfo {author} {\bibfnamefont {W.}~\bibnamefont {Zheng}}, \
  and\ \bibinfo {author} {\bibfnamefont {C.}~\bibnamefont {Clementi}},\
  }\href@noop {} {\bibfield  {journal} {\bibinfo  {journal} {Annu. Rev. Phys.
  Chem.}\ }\textbf {\bibinfo {volume} {64}},\ \bibinfo {pages} {295} (\bibinfo
  {year} {2013})}\BibitemShut {NoStop}%
\bibitem [{\citenamefont {Desiraju}(2011)}]{desi11ac}%
  \BibitemOpen
  \bibfield  {author} {\bibinfo {author} {\bibfnamefont {G.~R.}\ \bibnamefont
  {Desiraju}},\ }\href@noop {} {\bibfield  {journal} {\bibinfo  {journal}
  {Angew. Chem. Int. Ed. Engl.}\ }\textbf {\bibinfo {volume} {50}},\ \bibinfo
  {pages} {52} (\bibinfo {year} {2011})}\BibitemShut {NoStop}%
\bibitem [{\citenamefont {Arunan}\ \emph {et~al.}(2011)\citenamefont {Arunan},
  \citenamefont {Desiraju}, \citenamefont {Klein}, \citenamefont {Sadlej},
  \citenamefont {Scheiner}, \citenamefont {Alkorta}, \citenamefont {Clary},
  \citenamefont {Crabtree}, \citenamefont {Dannenberg},\ and\ \citenamefont
  {Hobza}}]{arun+11PureApplChem}%
  \BibitemOpen
  \bibfield  {author} {\bibinfo {author} {\bibfnamefont {E.}~\bibnamefont
  {Arunan}}, \bibinfo {author} {\bibfnamefont {G.~R.}\ \bibnamefont
  {Desiraju}}, \bibinfo {author} {\bibfnamefont {R.~A.}\ \bibnamefont {Klein}},
  \bibinfo {author} {\bibfnamefont {J.}~\bibnamefont {Sadlej}}, \bibinfo
  {author} {\bibfnamefont {S.}~\bibnamefont {Scheiner}}, \bibinfo {author}
  {\bibfnamefont {I.}~\bibnamefont {Alkorta}}, \bibinfo {author} {\bibfnamefont
  {D.~C.}\ \bibnamefont {Clary}}, \bibinfo {author} {\bibfnamefont {R.~H.}\
  \bibnamefont {Crabtree}}, \bibinfo {author} {\bibfnamefont {J.~J.}\
  \bibnamefont {Dannenberg}}, \ and\ \bibinfo {author} {\bibfnamefont
  {P.}~\bibnamefont {Hobza}},\ }\href@noop {} {\bibfield  {journal} {\bibinfo
  {journal} {Pure and applied chemistry}\ }\textbf {\bibinfo {volume} {83}},\
  \bibinfo {pages} {1637} (\bibinfo {year} {2011})}\BibitemShut {NoStop}%
\bibitem [{\citenamefont {Derewenda}\ \emph {et~al.}(1995)\citenamefont
  {Derewenda}, \citenamefont {Lee},\ and\ \citenamefont
  {Derewenda}}]{dere+95JMolBiol}%
  \BibitemOpen
  \bibfield  {author} {\bibinfo {author} {\bibfnamefont {Z.~S.}\ \bibnamefont
  {Derewenda}}, \bibinfo {author} {\bibfnamefont {L.}~\bibnamefont {Lee}}, \
  and\ \bibinfo {author} {\bibfnamefont {U.}~\bibnamefont {Derewenda}},\
  }\href@noop {} {\bibfield  {journal} {\bibinfo  {journal} {Journal of
  molecular biology}\ }\textbf {\bibinfo {volume} {252}},\ \bibinfo {pages}
  {248} (\bibinfo {year} {1995})}\BibitemShut {NoStop}%
\bibitem [{\citenamefont {Reed}\ \emph {et~al.}(1988)\citenamefont {Reed},
  \citenamefont {Curtiss},\ and\ \citenamefont {Weinhold}}]{reed+1988ChemRev}%
  \BibitemOpen
  \bibfield  {author} {\bibinfo {author} {\bibfnamefont {A.~E.}\ \bibnamefont
  {Reed}}, \bibinfo {author} {\bibfnamefont {L.~A.}\ \bibnamefont {Curtiss}}, \
  and\ \bibinfo {author} {\bibfnamefont {F.}~\bibnamefont {Weinhold}},\
  }\href@noop {} {\bibfield  {journal} {\bibinfo  {journal} {Chemical Reviews}\
  }\textbf {\bibinfo {volume} {88}},\ \bibinfo {pages} {899} (\bibinfo {year}
  {1988})}\BibitemShut {NoStop}%
\bibitem [{\citenamefont {Pend{\'a}s}\ \emph {et~al.}(2006)\citenamefont
  {Pend{\'a}s}, \citenamefont {Blanco},\ and\ \citenamefont
  {Francisco}}]{pend+2006JChemPhys}%
  \BibitemOpen
  \bibfield  {author} {\bibinfo {author} {\bibfnamefont {A.~M.}\ \bibnamefont
  {Pend{\'a}s}}, \bibinfo {author} {\bibfnamefont {M.}~\bibnamefont {Blanco}},
  \ and\ \bibinfo {author} {\bibfnamefont {E.}~\bibnamefont {Francisco}},\
  }\href@noop {} {\bibfield  {journal} {\bibinfo  {journal} {The Journal of
  chemical physics}\ }\textbf {\bibinfo {volume} {125}},\ \bibinfo {pages}
  {184112} (\bibinfo {year} {2006})}\BibitemShut {NoStop}%
\bibitem [{\citenamefont {Taylor}\ and\ \citenamefont
  {Kennard}(1984)}]{tayl-kenn84AccChemRes}%
  \BibitemOpen
  \bibfield  {author} {\bibinfo {author} {\bibfnamefont {R.}~\bibnamefont
  {Taylor}}\ and\ \bibinfo {author} {\bibfnamefont {O.}~\bibnamefont
  {Kennard}},\ }\href@noop {} {\bibfield  {journal} {\bibinfo  {journal}
  {Accounts of chemical research}\ }\textbf {\bibinfo {volume} {17}},\ \bibinfo
  {pages} {320} (\bibinfo {year} {1984})}\BibitemShut {NoStop}%
\bibitem [{\citenamefont {Matsumoto}(2007)}]{mats2007JChemPhys}%
  \BibitemOpen
  \bibfield  {author} {\bibinfo {author} {\bibfnamefont {M.}~\bibnamefont
  {Matsumoto}},\ }\href@noop {} {\bibfield  {journal} {\bibinfo  {journal} {The
  Journal of chemical physics}\ }\textbf {\bibinfo {volume} {126}},\ \bibinfo
  {pages} {054503} (\bibinfo {year} {2007})}\BibitemShut {NoStop}%
\bibitem [{\citenamefont {Kumar}\ \emph {et~al.}(2007)\citenamefont {Kumar},
  \citenamefont {Schmidt},\ and\ \citenamefont {Skinner}}]{kuma+07jcp}%
  \BibitemOpen
  \bibfield  {author} {\bibinfo {author} {\bibfnamefont {R.}~\bibnamefont
  {Kumar}}, \bibinfo {author} {\bibfnamefont {J.~R.}\ \bibnamefont {Schmidt}},
  \ and\ \bibinfo {author} {\bibfnamefont {J.~L.}\ \bibnamefont {Skinner}},\
  }\href@noop {} {\bibfield  {journal} {\bibinfo  {journal} {J. Chem. Phys.}\
  }\textbf {\bibinfo {volume} {126}},\ \bibinfo {pages} {204107} (\bibinfo
  {year} {2007})}\BibitemShut {NoStop}%
\bibitem [{\citenamefont {Jorgensen}\ \emph {et~al.}(1983)\citenamefont
  {Jorgensen}, \citenamefont {Chandrasekhar}, \citenamefont {Madura},
  \citenamefont {Impey},\ and\ \citenamefont {Klein}}]{jorg83JChemPhys}%
  \BibitemOpen
  \bibfield  {author} {\bibinfo {author} {\bibfnamefont {W.~L.}\ \bibnamefont
  {Jorgensen}}, \bibinfo {author} {\bibfnamefont {J.}~\bibnamefont
  {Chandrasekhar}}, \bibinfo {author} {\bibfnamefont {J.~D.}\ \bibnamefont
  {Madura}}, \bibinfo {author} {\bibfnamefont {R.~W.}\ \bibnamefont {Impey}}, \
  and\ \bibinfo {author} {\bibfnamefont {M.~L.}\ \bibnamefont {Klein}},\
  }\href@noop {} {\bibfield  {journal} {\bibinfo  {journal} {The Journal of
  chemical physics}\ }\textbf {\bibinfo {volume} {79}},\ \bibinfo {pages} {926}
  (\bibinfo {year} {1983})}\BibitemShut {NoStop}%
\bibitem [{\citenamefont {Plimpton}(1995)}]{plim95jcp}%
  \BibitemOpen
  \bibfield  {author} {\bibinfo {author} {\bibfnamefont {S.}~\bibnamefont
  {Plimpton}},\ }\href@noop {} {\bibfield  {journal} {\bibinfo  {journal} {J.
  Comput. Phys.}\ }\textbf {\bibinfo {volume} {117}},\ \bibinfo {pages} {1}
  (\bibinfo {year} {1995})}\BibitemShut {NoStop}%
\bibitem [{\citenamefont {Schneider}\ and\ \citenamefont
  {Stoll}(1978)}]{schn-stol78prb}%
  \BibitemOpen
  \bibfield  {author} {\bibinfo {author} {\bibfnamefont {T.}~\bibnamefont
  {Schneider}}\ and\ \bibinfo {author} {\bibfnamefont {E.}~\bibnamefont
  {Stoll}},\ }\href@noop {} {\bibfield  {journal} {\bibinfo  {journal} {Phys.
  Rev. B}\ }\textbf {\bibinfo {volume} {17}},\ \bibinfo {pages} {1302}
  (\bibinfo {year} {1978})}\BibitemShut {NoStop}%
\bibitem [{\citenamefont {Gonz{\'a}lez}\ and\ \citenamefont
  {Abascal}(2011)}]{gonz+11JChemPhys}%
  \BibitemOpen
  \bibfield  {author} {\bibinfo {author} {\bibfnamefont {M.~A.}\ \bibnamefont
  {Gonz{\'a}lez}}\ and\ \bibinfo {author} {\bibfnamefont {J.~L.~F.}\
  \bibnamefont {Abascal}},\ }\href {\doibase 10.1063/1.3663219} {\bibfield
  {journal} {\bibinfo  {journal} {The Journal of Chemical Physics}\ }\textbf
  {\bibinfo {volume} {135}},\ \bibinfo {pages} {224516} (\bibinfo {year}
  {2011})}\BibitemShut {NoStop}%
\bibitem [{\citenamefont {Luzar}\ and\ \citenamefont
  {Chandler}(1996)}]{luza-chan96nature}%
  \BibitemOpen
  \bibfield  {author} {\bibinfo {author} {\bibfnamefont {A.}~\bibnamefont
  {Luzar}}\ and\ \bibinfo {author} {\bibfnamefont {D.}~\bibnamefont
  {Chandler}},\ }\href@noop {} {\bibfield  {journal} {\bibinfo  {journal}
  {Nature}\ }\textbf {\bibinfo {volume} {379}},\ \bibinfo {pages} {55}
  (\bibinfo {year} {1996})}\BibitemShut {NoStop}%
\bibitem [{\citenamefont {Hayward}\ and\ \citenamefont
  {Reimers}(1997)}]{hayw-reim97jcp}%
  \BibitemOpen
  \bibfield  {author} {\bibinfo {author} {\bibfnamefont {J.~A.}\ \bibnamefont
  {Hayward}}\ and\ \bibinfo {author} {\bibfnamefont {J.~R.}\ \bibnamefont
  {Reimers}},\ }\href@noop {} {\bibfield  {journal} {\bibinfo  {journal} {J.
  Chem. Phys.}\ }\textbf {\bibinfo {volume} {106}},\ \bibinfo {pages} {1518}
  (\bibinfo {year} {1997})}\BibitemShut {NoStop}%
\bibitem [{\citenamefont {Bjerrum}(1952)}]{bjer52science}%
  \BibitemOpen
  \bibfield  {author} {\bibinfo {author} {\bibfnamefont {N.}~\bibnamefont
  {Bjerrum}},\ }\href@noop {} {\bibfield  {journal} {\bibinfo  {journal}
  {Science}\ }\textbf {\bibinfo {volume} {115}},\ \bibinfo {pages} {385}
  (\bibinfo {year} {1952})}\BibitemShut {NoStop}%
\bibitem [{\citenamefont {de~Koning}\ \emph {et~al.}(2006)\citenamefont
  {de~Koning}, \citenamefont {Antonelli}, \citenamefont {da~Silva},\ and\
  \citenamefont {Fazzio}}]{deko+2006prl}%
  \BibitemOpen
  \bibfield  {author} {\bibinfo {author} {\bibfnamefont {M.}~\bibnamefont
  {de~Koning}}, \bibinfo {author} {\bibfnamefont {A.}~\bibnamefont
  {Antonelli}}, \bibinfo {author} {\bibfnamefont {A.~J.}\ \bibnamefont
  {da~Silva}}, \ and\ \bibinfo {author} {\bibfnamefont {A.}~\bibnamefont
  {Fazzio}},\ }\href@noop {} {\bibfield  {journal} {\bibinfo  {journal}
  {Physical review letters}\ }\textbf {\bibinfo {volume} {96}},\ \bibinfo
  {pages} {075501} (\bibinfo {year} {2006})}\BibitemShut {NoStop}%
\bibitem [{\citenamefont {Ceriotti}\ and\ \citenamefont
  {Manolopoulos}(2012)}]{ceri-mano12prl}%
  \BibitemOpen
  \bibfield  {author} {\bibinfo {author} {\bibfnamefont {M.}~\bibnamefont
  {Ceriotti}}\ and\ \bibinfo {author} {\bibfnamefont {D.~E.}\ \bibnamefont
  {Manolopoulos}},\ }\href@noop {} {\bibfield  {journal} {\bibinfo  {journal}
  {Phys. Rev. Lett.}\ }\textbf {\bibinfo {volume} {109}},\ \bibinfo {pages}
  {100604} (\bibinfo {year} {2012})}\BibitemShut {NoStop}%
\bibitem [{\citenamefont {VandeVondele}\ \emph {et~al.}(2005)\citenamefont
  {VandeVondele}, \citenamefont {Krack}, \citenamefont {Mohamed}, \citenamefont
  {Parrinello}, \citenamefont {Chassaing},\ and\ \citenamefont
  {Hutter}}]{vand-krac05cpc}%
  \BibitemOpen
  \bibfield  {author} {\bibinfo {author} {\bibfnamefont {J.}~\bibnamefont
  {VandeVondele}}, \bibinfo {author} {\bibfnamefont {M.}~\bibnamefont {Krack}},
  \bibinfo {author} {\bibfnamefont {F.}~\bibnamefont {Mohamed}}, \bibinfo
  {author} {\bibfnamefont {M.}~\bibnamefont {Parrinello}}, \bibinfo {author}
  {\bibfnamefont {T.}~\bibnamefont {Chassaing}}, \ and\ \bibinfo {author}
  {\bibfnamefont {J.}~\bibnamefont {Hutter}},\ }\href@noop {} {\bibfield
  {journal} {\bibinfo  {journal} {Comput. Phys. Commun.}\ }\textbf {\bibinfo
  {volume} {167}},\ \bibinfo {pages} {103} (\bibinfo {year}
  {2005})}\BibitemShut {NoStop}%
\bibitem [{\citenamefont {Goedecker}\ \emph {et~al.}(1996)\citenamefont
  {Goedecker}, \citenamefont {Teter},\ and\ \citenamefont
  {Hutter}}]{goed+96prb}%
  \BibitemOpen
  \bibfield  {author} {\bibinfo {author} {\bibfnamefont {S.}~\bibnamefont
  {Goedecker}}, \bibinfo {author} {\bibfnamefont {M.}~\bibnamefont {Teter}}, \
  and\ \bibinfo {author} {\bibfnamefont {J.}~\bibnamefont {Hutter}},\
  }\href@noop {} {\bibfield  {journal} {\bibinfo  {journal} {Phys. Rev. B}\
  }\textbf {\bibinfo {volume} {54}},\ \bibinfo {pages} {1703} (\bibinfo {year}
  {1996})}\BibitemShut {NoStop}%
\bibitem [{\citenamefont {Becke}(1988)}]{beck88pra}%
  \BibitemOpen
  \bibfield  {author} {\bibinfo {author} {\bibfnamefont {A.~D.}\ \bibnamefont
  {Becke}},\ }\href@noop {} {\bibfield  {journal} {\bibinfo  {journal} {Phys.
  Rev. A}\ }\textbf {\bibinfo {volume} {38}},\ \bibinfo {pages} {3098}
  (\bibinfo {year} {1988})}\BibitemShut {NoStop}%
\bibitem [{\citenamefont {Lee}\ \emph {et~al.}(1988)\citenamefont {Lee},
  \citenamefont {Yang},\ and\ \citenamefont {Parr}}]{lee+88prb}%
  \BibitemOpen
  \bibfield  {author} {\bibinfo {author} {\bibfnamefont {C.}~\bibnamefont
  {Lee}}, \bibinfo {author} {\bibfnamefont {W.}~\bibnamefont {Yang}}, \ and\
  \bibinfo {author} {\bibfnamefont {R.~G.}\ \bibnamefont {Parr}},\ }\href@noop
  {} {\bibfield  {journal} {\bibinfo  {journal} {Phys. Rev. B}\ }\textbf
  {\bibinfo {volume} {37}},\ \bibinfo {pages} {785} (\bibinfo {year}
  {1988})}\BibitemShut {NoStop}%
\bibitem [{\citenamefont {Ceriotti}(2010)}]{GLE4MD}%
  \BibitemOpen
  \bibfield  {author} {\bibinfo {author} {\bibfnamefont {M.}~\bibnamefont
  {Ceriotti}},\ }\href@noop {} {\enquote {\bibinfo {title} {{GLE4MD}},}\
  }\bibinfo {howpublished} {http://epfl-cosmo.github.io/gle4md} (\bibinfo
  {year} {2010})\BibitemShut {NoStop}%
\bibitem [{\citenamefont {Ceriotti}\ \emph {et~al.}(2014)\citenamefont
  {Ceriotti}, \citenamefont {More},\ and\ \citenamefont
  {Manolopoulos}}]{ceri+14cpc}%
  \BibitemOpen
  \bibfield  {author} {\bibinfo {author} {\bibfnamefont {M.}~\bibnamefont
  {Ceriotti}}, \bibinfo {author} {\bibfnamefont {J.}~\bibnamefont {More}}, \
  and\ \bibinfo {author} {\bibfnamefont {D.~E.}\ \bibnamefont {Manolopoulos}},\
  }\href@noop {} {\bibfield  {journal} {\bibinfo  {journal} {Comput. Phys.
  Commun.}\ }\textbf {\bibinfo {volume} {185}},\ \bibinfo {pages} {1019}
  (\bibinfo {year} {2014})}\BibitemShut {NoStop}%
\bibitem [{\citenamefont {Ceriotti}\ \emph {et~al.}(2013)\citenamefont
  {Ceriotti}, \citenamefont {Cuny}, \citenamefont {Parrinello},\ and\
  \citenamefont {Manolopoulos}}]{ceri+13pnas}%
  \BibitemOpen
  \bibfield  {author} {\bibinfo {author} {\bibfnamefont {M.}~\bibnamefont
  {Ceriotti}}, \bibinfo {author} {\bibfnamefont {J.}~\bibnamefont {Cuny}},
  \bibinfo {author} {\bibfnamefont {M.}~\bibnamefont {Parrinello}}, \ and\
  \bibinfo {author} {\bibfnamefont {D.~E.}\ \bibnamefont {Manolopoulos}},\
  }\href@noop {} {\bibfield  {journal} {\bibinfo  {journal} {Proc. Natl. Acad.
  Sci. USA U. S. A.}\ }\textbf {\bibinfo {volume} {110}},\ \bibinfo {pages}
  {15591} (\bibinfo {year} {2013})}\BibitemShut {NoStop}%
\bibitem [{\citenamefont {More}\ and\ \citenamefont
  {Manolopoulos}(2014)}]{nh3unpub}%
  \BibitemOpen
  \bibfield  {author} {\bibinfo {author} {\bibfnamefont {J.}~\bibnamefont
  {More}}\ and\ \bibinfo {author} {\bibfnamefont {D.}~\bibnamefont
  {Manolopoulos}},\ }\href@noop {} {}\bibinfo {howpublished} {private
  communication} (\bibinfo {year} {2014})\BibitemShut {NoStop}%
\bibitem [{\citenamefont {Baroni}\ \emph {et~al.}()\citenamefont {Baroni},
  \citenamefont {{Dal Corso}}, \citenamefont {de~Gironcoli}, \citenamefont
  {Giannozzi}, \citenamefont {Cavazzoni}, \citenamefont {Ballabio},
  \citenamefont {Scandolo}, \citenamefont {Chiarotti}, \citenamefont {Focher},
  \citenamefont {Pasquarello}, \citenamefont {Laasonen}, \citenamefont {Trave},
  \citenamefont {Car}, \citenamefont {Marzari},\ and\ \citenamefont
  {Kokalj}}]{PWSCF}%
  \BibitemOpen
  \bibfield  {author} {\bibinfo {author} {\bibfnamefont {S.}~\bibnamefont
  {Baroni}}, \bibinfo {author} {\bibfnamefont {A.}~\bibnamefont {{Dal Corso}}},
  \bibinfo {author} {\bibfnamefont {S.}~\bibnamefont {de~Gironcoli}}, \bibinfo
  {author} {\bibfnamefont {P.}~\bibnamefont {Giannozzi}}, \bibinfo {author}
  {\bibfnamefont {C.}~\bibnamefont {Cavazzoni}}, \bibinfo {author}
  {\bibfnamefont {G.}~\bibnamefont {Ballabio}}, \bibinfo {author}
  {\bibfnamefont {S.}~\bibnamefont {Scandolo}}, \bibinfo {author}
  {\bibfnamefont {G.}~\bibnamefont {Chiarotti}}, \bibinfo {author}
  {\bibfnamefont {P.}~\bibnamefont {Focher}}, \bibinfo {author} {\bibfnamefont
  {A.}~\bibnamefont {Pasquarello}}, \bibinfo {author} {\bibfnamefont
  {K.}~\bibnamefont {Laasonen}}, \bibinfo {author} {\bibfnamefont
  {A.}~\bibnamefont {Trave}}, \bibinfo {author} {\bibfnamefont
  {R.}~\bibnamefont {Car}}, \bibinfo {author} {\bibfnamefont {N.}~\bibnamefont
  {Marzari}}, \ and\ \bibinfo {author} {\bibfnamefont {A.}~\bibnamefont
  {Kokalj}},\ }\href@noop {} {\enquote {\bibinfo {title} {{PWSCF}},}\
  }\BibitemShut {NoStop}%
\bibitem [{\citenamefont {Perdew}\ \emph {et~al.}(1996)\citenamefont {Perdew},
  \citenamefont {Burke},\ and\ \citenamefont {Ernzerhof}}]{perd+96prl}%
  \BibitemOpen
  \bibfield  {author} {\bibinfo {author} {\bibfnamefont {J.~P.}\ \bibnamefont
  {Perdew}}, \bibinfo {author} {\bibfnamefont {K.}~\bibnamefont {Burke}}, \
  and\ \bibinfo {author} {\bibfnamefont {M.}~\bibnamefont {Ernzerhof}},\
  }\href@noop {} {\bibfield  {journal} {\bibinfo  {journal} {Phys. Rev. Lett.}\
  }\bibinfo {series} {Phys. Rev. Lett. (USA)},\ \textbf {\bibinfo {volume}
  {77}},\ \bibinfo {pages} {3865} (\bibinfo {year} {1996})}\BibitemShut
  {NoStop}%
\bibitem [{\citenamefont {Vanderbilt}(1990)}]{vand90prb}%
  \BibitemOpen
  \bibfield  {author} {\bibinfo {author} {\bibfnamefont {D.}~\bibnamefont
  {Vanderbilt}},\ }\href@noop {} {\bibfield  {journal} {\bibinfo  {journal}
  {Phys. Rev. B}\ }\textbf {\bibinfo {volume} {41}},\ \bibinfo {pages} {7892}
  (\bibinfo {year} {1990})}\BibitemShut {NoStop}%
\end{thebibliography}
\end{document}